\documentclass{aa}

\usepackage{graphicx}
\usepackage{txfonts}
\usepackage{hyperref}
\hypersetup{
    colorlinks=true
    linkcolor=blue,
    filecolor=blue,      
    urlcolor=cyan,
  }

\usepackage{natbib,twoopt}
\newcommand{\alfa}{$\alpha$}
\newcommand{\AF}{[\alfa/Fe]}

\newcommand{\SNR}{$S/N$}

\newcommand{\CeII}{Ce~{\sc ii}}

\newcommand{\Gaia}{{\it Gaia}}
\newcommand{\flags}{{\it flags\_gspspec}}
\newcommand{\gspspec}{{GSP-Spec}}
\newcommand{\T}{$T_{\rm eff}$}
\newcommand{\g}{log($g$)}

\newcommand{\meta}{[M/H]}
\newcommand{\alphaFe}{[$\alpha$/Fe]}

\newcommand{\CeFe}{[Ce/Fe]}
\newcommand{\CaFe}{[Ca/Fe]}

\newcommand{\MgFe}{[Mg/Fe]}
\newcommand{\CeM}{[Ce/M]}

\usepackage{listings}

\begin{document} 

   \title{The cerium content of the Milky Way as revealed by \Gaia~DR3 \gspspec\ abundances}


   \author{G. Contursi\inst{1} 
          \and
          P. de Laverny\inst{1} 
          \and
          A. Recio-Blanco\inst{1} 
         \and
         E.Spitoni\inst{1} 
         \and
         P. A. Palicio\inst{1} 
         \and
        E. Poggio\inst{1} 
        \and
         V. Grisoni\inst{2,3} 
         \and
         G. Cescutti\inst{4,5,6}
         \and
         F. Matteucci\inst{4,5,6} 
         \and
        L. Spin\inst{7} 
        \and
        M.A. \'Alvarez\inst{8}
        \and
        G. Kordopatis\inst{1} 
        \and
        C. Ordenovic \inst{1}
        \and
        I.~ Oreshina-Slezak \inst{1}
        \and
        H. Zhao\inst{1,9}
          }

   \institute{Université Côte d'Azur, Observatoire de la Côte d'Azur, CNRS, Laboratoire Lagrange, Bd de l'Observatoire, CS 34229, 06304 Nice cedex 4, France
   \and
Dipartimento di Fisica e Astronomia, Università degli Studi di Bologna, Via Gobetti 93/2, I-40129 Bologna, Italy
\and
INAF, Osservatorio di Astrofisica e Scienza dello Spazio, Via Gobetti 93/3, 40129 Bologna, Italy
  \and 
  Dipartimento di Fisica, Sezione di Astronomia, Università di Trieste, via G.B. Tiepolo 11, I-34131, Trieste, Italy
  \and
INAF Osservatorio Astronomico di Trieste, via G.B. Tiepolo 11, I-34131, Trieste, Italy
  \and
INFN Sezione di Trieste, via Valerio 2, 34134 Trieste, Italy
\and
INAF - Osservatorio astronomico di Padova, Vicolo Osservatorio
5, 35122 Padova, Italy
\and
CIGUS CITIC - Department of Computer Science and Information Technologies, University of A Coru\~{n}a, Campus de Elvi\~{n}a s/n, A Coru\~{n}a, 15071, Spain
\and
          Purple Mountain Observatory, Chinese Academy of Sciences, Nanjing 210023, PR China
  }

   \date{Received July 2022; accepted December 2022}

 
  \abstract
   {The recent \Gaia\ third data release contains a homogeneous analysis of millions of high-quality {\it Radial Velocity Spectrometer} (RVS) stellar spectra by the \gspspec\ module. This led to the estimation of millions of individual chemical abundances and allows us to chemically map the Milky Way. The published \gspspec\ abundances include three heavy elements produced by neutron-captures in stellar interiors: Ce, Zr, and Nd.}
   {We study the Galactic content in cerium based on these \Gaia/RVS data and discuss the chemical evolution of this element.
   }
   {We used a sample of about 30,000 local thermal equilibrium Ce abundances, selected after applying different combinations of \gspspec~flags. Based on the \Gaia\ DR3 astrometric data and radial velocities,  we explore the cerium content in the Milky Way and, in particular, in its halo and disc components.}
   {
   The high quality of the Ce \gspspec\ abundances is quantified through literature comparisons.
   We found a rather flat \CeFe\ versus \meta\ trend. We also found a flat radial gradient in the disc derived from field stars and, independently,  from about 50 open clusters. This agrees with previous studies. The \CeFe\ vertical gradient was also estimated.
   We also report an increasing [Ce/Ca] versus [Ca/H] in the disc, illustrating the late contribution of asymptotic giant branch stars with respect to supernovae of type~II. Our cerium abundances in the disc, including the young massive population, are well reproduced by a new {\it three-infall} chemical evolution model.  
   In the halo population, the M~4 globular cluster is found to be enriched in cerium. Moreover, 11 stars with cerium abundances belonging to the Thamnos, Helmi Stream, and Gaia-Sausage-Enceladus accreted systems were identified from chemo-dynamical diagnostics. We found that the Helmi Stream might be slightly underabundant in cerium compared to the two other systems.}
   {This work illustrates the high quality of the \gspspec\ chemical abundances, which significantly contribute to unveiling the heavy-element evolution history of the Milky Way.}

   \keywords{Galaxy: abundances, Stars: abundances, Galaxy: evolution, Galaxy: disc, Galaxy: halo, surveys
               }

   \maketitle
%

\section{Introduction}


Our understanding of the Milky Way has made a great leap forwards through the different data releases of the Gaia mission. 
The third release \citep{Vallenari22} consists of a major and unique step because it includes a large variety of new data products, including, in particular, an extensive characterisation of the \Gaia\ sources. In this context, the module called general stellar parametrizer from spectroscopy \citep[\gspspec\  hereafter; see][]{PVP_Ale} has estimated atmospheric parameters (effective temperature \T, surface gravity \g, global metallicity \meta, and abundances of $\alpha$-elements with respect to iron \AF) as well as individual chemical abundances of up to a dozen  elements\footnote{see \url{https://www.cosmos.esa.int/web/gaia/iow_20210709}. {These elements are N, Mg, Si, S, Ca, Ti, Fe I, Fe II, Ni, Zr, Ce, and Nd.}}  for about 5.6 million stars that have been observed by the Radial Velocity Spectrometer  \cite[RVS hereafter;][]{RVS18, Katz22}.

Three of these 13 chemical elements are produced by neutron capture in the inner layers of some specific stages of stellar evolution: zirconium (Z = 40), cerium (Z = 58), and neodynium (Z = 60). 
According to the seminal work of \citet{Burbidge57}, neutron capture occurs through two main processes: the rapid ($r$-) and slow ($s$-) processes (slow and rapid referring to the timescale of the neutron captures with respect to the $\beta$-decay). The latter takes place in lower neutron densities and on longer timescales than the $r$-process. 
The main formation sites of the $r$-process elements are
still under discussion: merging of neutron stars
or of neutron star - black hole binary systems \citep[respectively]{Freiburghaus99, Surman08}, neutrino-induced winds from the core-collapse of supernovae \citep{Woosley94}, and/or rotating polar jets from core-collapse supernovae \citep{Nishimura06}.

The formation sites of the $s$-process, in contrast, are better understood. The distribution of solar abundances shows  
three peaks located around the atomic mass numbers $A$=90, 138, and 208. Sr, Y, and Zr represent the first peak, Ba, La, and Ce the second peak, and Pb the third peak. Even though all these elements are mainly formed via s-process \citep{Nikos18}, their formation sites can differ. The $s$-process can be decomposed into three sub-processes, each of which populates a different peak \citep[see][and references therein]{Kappeler89}. 
First, massive stars ($\gtrsim$ 8-10~M$_\odot$) preferentially cause the so-called weak process (especially producing first peak elements such as Zr), where neutrons are mainly provided by the $^{22}$Ne($\alpha$, n)$^{25}$Mg reaction in the convective He-burning core and C-burning shell \citep{Peters68, Lamb77, Pignatari10}. On the other hand, 
low- and intermediate-mass Asymptotic giant branch stars {at solar metallicity} 
produce the so-called $main$ $s$-process such as Ce and Nd through neutrons that are mainly produced{} by the $^{13}$C($\alpha$,n)$^{16}$O reaction \citep{Arlandini99, Busso99, KL14, Bisterzo11, Bisterzo15}. This reaction takes place in the so-called $^{13}$C-pocket, between the H- and He-burning shells. This $^{13}$C pocket is formed through a sequence reaction $^{12}$C(p,$\gamma$)$^{13}$N($\beta^+$)$^{13}$C through the partial mixing of protons from the convective H-rich envelope  into the $^{12}$C region during the third dredge-up. The $^{22}$Ne($\alpha$, n)$^{25}$Mg reaction also contributes to the convective thermal pulse. Moreover, we note that rotating massive stars at low metallicity ([Fe/H] $<$ -0.5~dex) seem to have an impact on the $main$ $s-$ process elements \citep{Nikos18}. Finally, low-metallicity low-mass AGB can produce elements of the third peak (e.g. Pb) through the $strong$ $s$-process \citep{CR67, Gallino98, Travaglio01}

However, this rather simple picture is blurred by the fact that neutron-capture elements can be produced by a combination of the $s$- and $r$- processes. For instance, \citet{Arlandini99} found that at the epoch of the Solar System formation, cerium could come at a level of 77\% from $s$-process production. This was later confirmed by \citet{Bisterzo16} and \citet{Nikos18}, who report  a $s$-process contribution of about 80\%.

For all these reasons, studying the Ce content in the Milky Way allows us to probe its different production sites and, in particular, the main $s$- process. 
Several studies of neutron-capture elements and more especially, of cerium abundances in the Galactic disc,
can be found in the literature, for instance, \citet{Reddy06, Mishenina13, BB16, DM17, Rebecca19, Griffith21} for field star studies, and \citet{Magrini18, Spina21, SalesSilva22} for open clusters. 
Global flat trends of \CeFe\ versus \meta\ were found in most works, even though some report a small decreasing trend for high [Fe/H] values \citep{Mashonkina07}. This trend agrees with chemical models \citep{Nikos18,  grisoni2020}. Moreover, a correlation between cerium abundances and ages has been proposed based on open clusters \citep{SalesSilva22}, {where \CeFe~appears to decrease with age up to 8 Gyr. A similar trend was obtained for field stars \citep{BB16}. They additionally found an increase in \CeFe\ with ages older than 8 Gyr.}

With the recent delivery of the third $Gaia$ data release, these studies can be extended to a much larger stellar sample. The aim of the present paper is therefore to complement the first description of Ce based on the \Gaia\ \gspspec\ abundances that was
presented by \citet{PVP_Ale}. 
This work is composed as follows. Section~2 describes the selection of the sample of Ce abundances, and Sect.~3 presents its spatial, chemical, and dynamical properties.  Sect.~4 presents the Galactic disc content in cerium with the derivation of radial and vertical gradients, the comparison with a new chemical evolution model, and the Ce content of open clusters. Then, Sect.~5 explores Ce abundances in the Galactic halo and, in particular, in the M~4 globular cluster and in accreted dwarf galaxies and stellar streams. Finally, the conclusions of this work are presented in Sect.~6.

\section{Sample stars of cerium abundances}
\label{Sect:Sample}

This section presents the recently published  \Gaia\ DR3 cerium abundances. We select the best working samples for further analysis (see Appendix~\ref{Append} for the corresponding catalogue queries). 

\subsection{\Gaia\ \gspspec\ local thermal equilibrium cerium abundances}
First, we briefly recall that the chemical analysis of the \Gaia-RVS spectra (R $\sim$ 11,500) was performed by the \gspspec\ module \citep{GSPspecDR3}
through the GAUGUIN algorithm \citep{2012ada..confE...2B, RB16} and a specific grid of synthetic spectra covering the whole stellar atmospheric parameters space and with varying Ce abundances. {Briefly, to derive chemical abundances, GAUGUIN builds a reference grid spectrum (1D) from this large 5D cerium grid that is interpolated at the atmospheric parameters of the analysed star. These atmospheric parameters are provided by the MatisseGauguin method (see Sect.~6 of \citet{GSPspecDR3} for more details). The observed spectrum of the analysed star is normalised, and a second normalisation around the line is then performed to readjust the continuum locally \citep{Pablo20, GSPspecDR3}. Finally, the minimum of the quadratic distance between the observed and reference spectra is computed in a wavelength range close to the line. This provides the initial guess of the Ce abundance, from which the Gauss-Newton algorithm obtains the final abundance. The second normalisation and abundance windows from which the cerium abundances are derived are provided in Table B.1. of \citet{GSPspecDR3}. }
The derived local thermodynamical equilibrium (LTE) abundances rely on the line data
of \citet{BestArticleEver} and assume  \citet{Grevesse07} 
solar abundances.

Cerium abundances are determined from a triplet of \CeII~lines 
centred around 851.375 nm (in vacuum). {Table~\ref{Tab:Lines} reports the atomic data of this triplet that we adopted for the analysis: air and vacuum wavelengths\footnote{The conversion between air and vacuum wavelength was made following \citet{1994Metro..31..315B}.}, excitation energy of the lower level, and oscillator strength. These atomic data were not calibrated astrophysically. Moreover, for some specific combinations of atmospheric parameters, this cerium triplet might be slightly blended with a weak CN line whose central wavelength is about 851.25~nm.} 

\begin{figure}[t!]
        \centering
        \includegraphics[scale = 0.15]{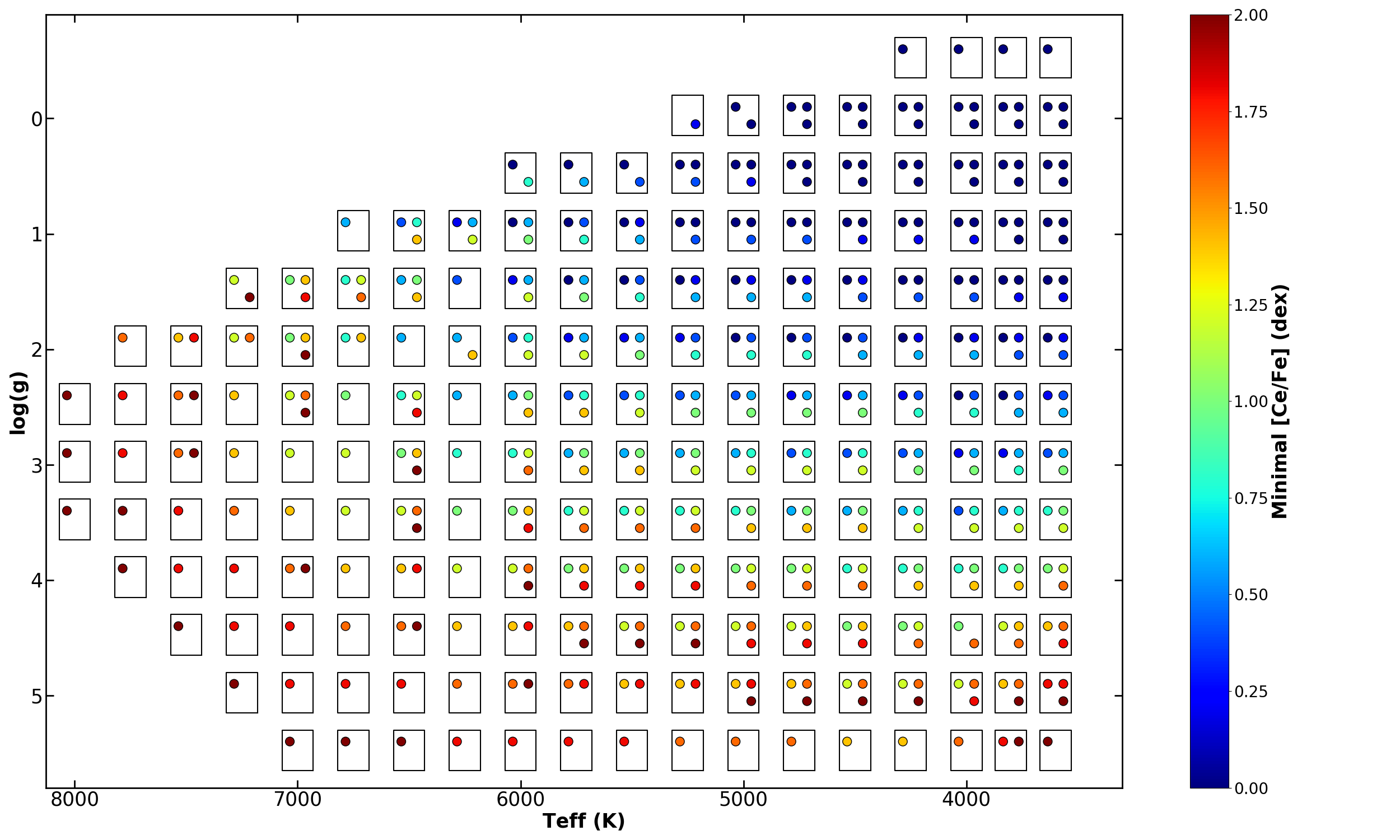}
        \caption{{Kiel diagram colour-coded with the lowest cerium abundance (in dex) that could be detected in a spectrum whose 
        (normalised) line core flux is 0.5\% deeper than that of a reference spectrum with \CeFe~= -2.0 dex. For each combination of effective temperature and surface gravity, we estimated this lowest cerium abundance for three values of [M/H]: 0.0, -0.5, and -1.0~dex (from top to bottom and left to right in each small square).}}
        \label{fig:CeKiel}
\end{figure}

An example of this cerium feature is provided in Fig.~9 of \citet{GSPspecDR3}. 
{The detectability of the cerium triplet is illustrated in Fig.~\ref{fig:CeKiel} , which presents a Kiel diagram colour-coded with the minimum cerium abundance (in dex) that could be measured for metallicities varying between -1.0 to 0.0~dex. This was estimated from the grid of synthetic spectra in which
we searched for the Ce abundance corresponding to a (normalised) flux decrease of 0.5\% at the Ce line core with respect to a reference spectrum with \CeFe~= -2.0 dex (i.e. no Ce and the lowest cerium abundance in the grid of reference spectra). The cerium lines are more easily detectable in AGB and RGB stars (\CeFe>0 dex), whereas higher cerium abundances are required for a possible detection in dwarf stars. This may lead to the observational biases that are discussed in Sect. \ref{Sect:Final-chemical} and Sect. \ref{Sect:Bias}.}

We recall that according to \citet{Lawler09}, cerium has four main stable isotopes, and two of them are dominant (88.45\% for $^{140}$Ce and 11.11\% for $^{142}$Ce). Then, 0.19\% and 0.25\% of Ce are found in $^{146}$Ce and $^{138}$Ce, respectively.

        \begin{table}[ht]
                \centering
                \caption{{\label{Tab:Lines} Adopted cerium line data.}}
                \begin{tabular}{lcccc}
                        \hline
                        \hline
                        Element & $\lambda_{air}$ & $\lambda_{vac}$& E & log(\textit{gf}) \\
                         & (nm) & (nm) & (eV) & \\
                         \hline
                         Ce II & 851.1337 & 851.3676 & 0.357 & -2.530 \\
                         Ce II & 851.1473 & 851.3812 & 2.004 & -2.120 \\
                         Ce II & 851.1521 & 851.3859 & 0.328 & -2.840 \\
                         \hline
                \end{tabular}
        \end{table}


Together with other chemical abundances, \gspspec\ provides several quality flags that are recommended for selecting the best data. These flags are related to several effects that could affect the stellar parametrisation, for instance, possible biases induced by radial velocity uncertainties, rotational broadening, or flux noise.
Moreover, two flags specifically refer to the determination of the cerium chemical abundance (Ce$UpLim$ and Ce$Uncer$).
The Ce$UpLim$ flag is an indicator of the line depth with respect to noise level, which corresponds to the detectability limit defined as the upper limit. The smaller this flag, the better the measurement. The Ce$Uncer$ flag is defined as the reliability of the abundance uncertainty considering the atmospheric parameters and \SNR. We refer to  \citet{GSPspecDR3} for a complete definition of these flags.

Out of the $\sim$5.5 million stars parametrised by \gspspec, 103,948 have a cerium abundance, without considering any flag restriction (we refer to this entire Ce catalogue as the {\it complete sample} hereafter). As this study aims to describe the largest possible sample with the most accurate Ce measurements, we applied some flag restrictions to define our working samples and compare the \gspspec\ \CeFe\ values with the literature values. 

\subsection{Comparison catalogues of cerium abundances}
In order to validate the \gspspec\ cerium abundances, they were compared
to the abundances of APOGEE-DR17 \citep{APOGEE17}, \citet{Rebecca19} (APOGEE and F19 hereafter, respectively), and GALAH DR3 \citep{GALAHDR3}.

{We first note that there are no GALAH stars in common with F19 after the recommended GALAH flag values were applied ($snr\_c3\_iraf$ > 30, $flag\_sp$ == 0, $flag\_fe\_h$ == 0 and $flag\_Ce\_fe$ == 0). Nevertheless, we compare below our \gspspec~cerium abundances with the abundance from GALAH in Sect. \ref{Sect:Comp}. These GALAH Ce abundances were derived from one cerium line around 477.3941 nm (air) from spectra with R$\sim$28,000. A zero-point calibration was applied. Using these recommended flags, we found 278,163 GALAH cerium abundances.}

The F19 cerium abundance sample is composed of 336 stars observed at highweresolution (R $\sim$ 67 000). Their abundances are obtained from a \CeII~line located at 604.3373 nm in the air, adopting 
solar composition from \citet{Grevesse15}.

APOGEE cerium abundances were derived from spectra with R$\sim$22,500 and \SNR$>$100, using multiple cerium lines. For our comparison purpose, we selected the best APOGEE non-calibrated Ce abundances. We then filtered all stars with APOGEE ANDFLAGs, ASPCAPFLAGs, RV\_FLAGs, and STARFLAGs $\neq$ 0. We also removed stars with a non-zero third binary digit of the EXTRATARG flag and a non-zero value for the sixth and twenty-sixth binary digits of each member of PARAMFLAG tuple. Finally, we kept only APOGEE cerium abundances with an uncertainty smaller than 0.2~dex and found 53,310 stars. {Our flag selection is the recommended optimized version of APOGEE flags.}
The F19 and APOGEE samples can be compared between each other. We found 32 stars with high-quality Ce abundances in common. They have a mean difference of -0.14 dex, in the sense APOGEE minus F19, indicating that the two studies are probably not at the same reference level. {Calibrated APOGEE abundances lead to a larger difference of -0.20~dex. We therefore consider only non-calibrated APOGEE abundances in the following.} F19 Ce abundances are indeed almost always higher than those from APOGEE. This difference is even larger for some APOGEE cerium-poor stars that are found to be $\sim$0.3-0.4~dex more enriched in Ce by F19, although no significant differences in atmospheric parameters are present. Nevertheless, the standard deviation of the Ce abundance differences in the two samples is equal to 0.13 dex, revealing a rather good agreement between the two studies. 
We note that similar systematic differences (different reference scales) of cerium abundances at solar metallicity have been reported by F19 when they compared their own abundances with those of \citet{BB16}.

\begin{table*}[t]
        \centering
        
        \begin{tabular}{lccccccccc}
                \hline
                \hline
                 &  \multicolumn{3}{c}{\it Strict Select.} & \multicolumn{3}{c}{\it Low-Uncer. Samp.} & \multicolumn{3}{c}{\it Complete Samp.}\\
                 & F19 & APO & GLH & F19 & APO & GLH & F19 & APO & GLH\\
                \hline
                mean     & 0.03 & -0.06 & - & 0.00 & -0.16 & -0.31 & 0.00  & -0.27 & -0.44\\
                std      & 0.05 & 0.09  & - & 0.15 & 0.25  & 0.20 & 0.15 & 0.30 & 0.32\\
                N$_{comp}$  & 9  & 2  & 0  & 105  & 101 & 44  & 122  & 187 & 333\\
                N$_{Ce}$ &   \multicolumn{3}{c}{493} & \multicolumn{3}{c} {29,991} & \multicolumn{3}{c}{103,948}\\
                \hline
        \end{tabular}
                \caption{\label{Tab:Comp} Comparison (mean and standard deviation of the differences) between \gspspec\ cerium abundances and those of F19, APOGEE, {and GALAH} for different samples of Ce \gspspec\ abundances. {The \CeFe~mean values are computed in the sense literature minus \gspspec. }N$_{comp}$ indicates the number of stars found for the comparison, and N$_{Ce}$ is the total number
                of \gspspec\ stars with Ce abundances when the corresponding flag selection was applied (see text for more details).}
\end{table*}

\subsection{Definition of the {\it low-uncertainty sample} of \gspspec\ Ce abundances}
We then compare in Tab.~\ref{Tab:Comp} the differences between \gspspec\ cerium abundances and those of F19, APOGEE, {and GALAH} for three possible \flags~selections. We also indicate the number of stars (N$_{col}$) used for the comparison, as well as the total number of selected \gspspec~stars (N$_{Ce}$) when the considered flags were applied.

We first show  this comparison in the left column by adopting stars for which (i) all their \flags=0 (including those related to Ce abundances), (ii) Ce uncertainties smaller than 0.2~dex \cite[estimated from Monte Carlo simulations; see][]{GSPspecDR3},
(iii) $vbroad \le$ 13 km/s (since deriving accurate chemical abundances 
can become difficult for fast-rotating stars)\footnote{This value is a good compromise between keeping a high number of stars and good-quality abundances.} , (iv)  \T~$\le$ 5400~K (tests with synthetic spectra show that the Ce lines become too weak to be detected in {hotter} star spectra), and (v) \g~$\le$3.5 (tests with synthetic spectra revealed that the Ce line becomes difficult to analyse in dwarf stars)\footnote{Nevertheless, we found a dwarf star (ID=5373254711531881728, \g~= 4.21, \T~= 4775.0 K , \meta~=-0.24 dex, \SNR~=81) that appears to be strongly enriched in cerium. (\CeFe$\ge$2.0 dex, 2.0 being the Ce reference grid high-border value). {No clear sign of binarity has been found in the \Gaia\ astrometric data for this star, which, moreover, does not belong to the \Gaia\ binary catalog. No abundance of other heavy elements has been found in the literature for these stars either.}}
. We finally found 493 stars that satisfied these criteria (referred to as the {\it strict selection} hereafter).
Among them, we found only 9, 2 {and 0} stars in common with F19, APOGEE {and GALAH}, respectively. An excellent agreement between \CeFe\ \gspspec, F19 and APOGEE values is found.

For comparison purpose, we provide a similar comparison in the right column of Tab.~\ref{Tab:Comp}, but considering the {\it complete sample} of Ce abundances. 
The number of stars in common is much larger, and the agreement between \gspspec\ and F19 is still very good (no bias and dispersion equal to 0.15~dex). In contrast, the agreement with APOGEE {and GALAH} is worse. The large bias can be explained by the fact that \gspspec,  APOGEE, {and GALAH} Ce abundances are not on the same scale (as is also the case for F19 and APOGEE, as mentioned in the previous subsection). {The dispersion is also larger, maybe because the some of APOGEE stars found in \gspspec\ are fainter than those in F19.}

From these considerations and in order to select a large enough but still accurate sample of Ce abundances, we 
defined a specific combination of the \gspspec\ flags by relaxing the $extrapol$ and $KMgiantPar$ flags together with those specifically related to Ce abundances. Briefly, the $extrapol$ flag indicates if the \gspspec\ atmospheric parameters are extrapolated beyond the limits of the reference grid and $KMgiantPar$ refers to extremely cool giant stars whose \T\ and \g\ have been corrected and set to given specific values because of parametrisation issues. We again refer to \citet{GSPspecDR3} for a complete definition of these flags.
Our best combination was found by adopting $CeUpLim \le 2$, $CeUncer \le 1$ and $extrapol\le 1$.
We also fixed $KMgiantPar \le 1$, which is associated with a {\it goodness of fit} \citep[referred as $gof$ hereafter; see][for its definition]{GSPspecDR3} lower than
-3.75. All the other flags were set to 0 to ensure a good stellar parametrisation. As for the {\it strict selection} sample, we also only selected stars with $vbroad \le$ 13 km/s, Ce uncertainties $\le$ 0.2 dex, \T $\le$ 5400~K, and \g $<$ 3.5. This {\it low-uncertainty sample} contains 29,991 stars with accurate measurements of the Ce abundances. 

\subsection{Comparison of the {\it low-uncertainty sample} with the reference catalogues}
\label{Sect:Comp}
\begin{figure}[t!]
        \centering
        \includegraphics[scale = 0.15]{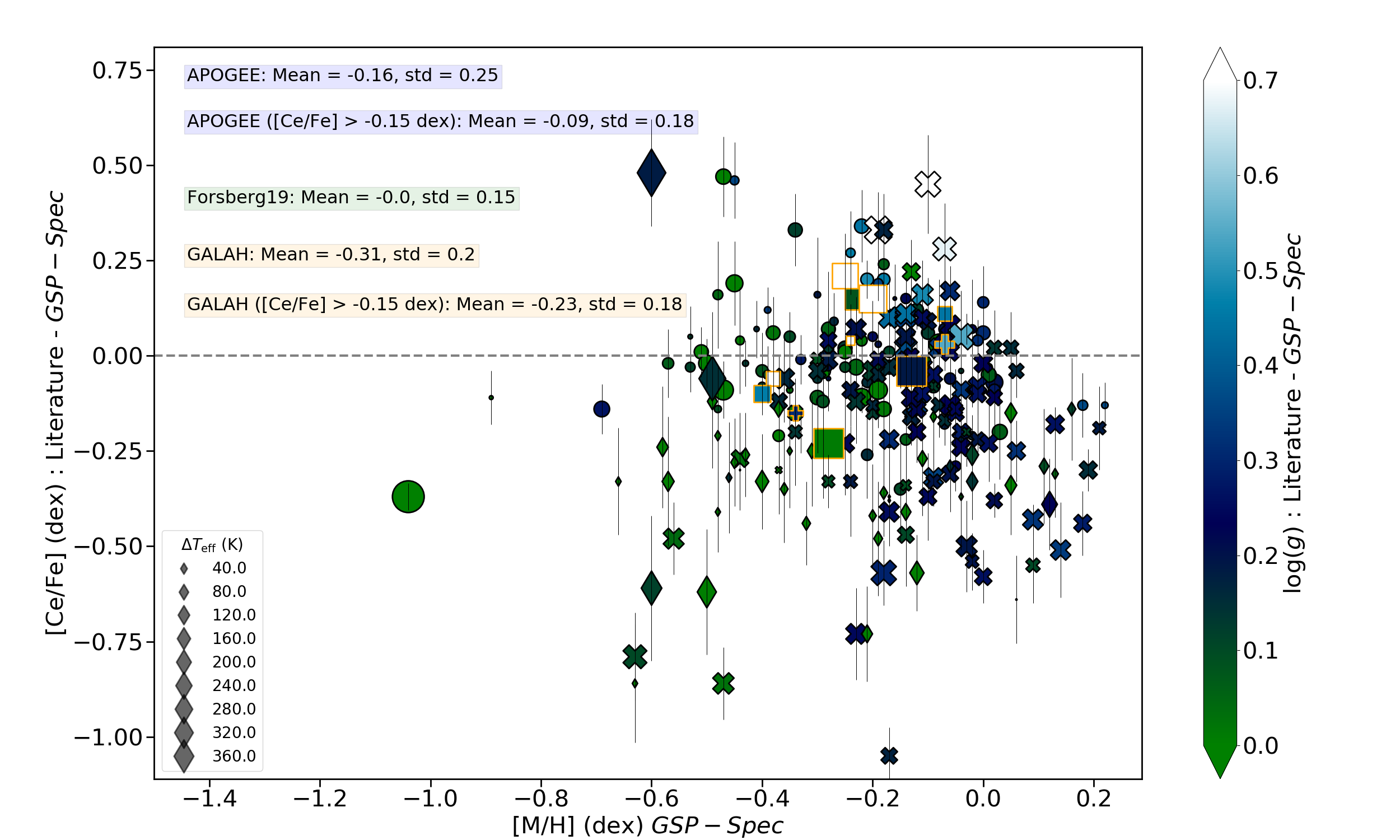}
        \caption{Comparison between \gspspec\ {\it low-uncertainty sample} cerium abundance and  literature values as a function of the \gspspec\ metallicity (crosses refer to APOGEE data, circles to F19, and {diamonds to GALAH. Squares and plus markers refer to the nine and two stars in common between the strict sample and F19 and APOGEE, respectively}). The points are colour-coded with the difference in \g,\ and their size is proportional to differences in \T. The vertical error bars indicate the uncertainty of \gspspec~Ce abundance. The mean and the standard deviation of the cerium abundance differences between \gspspec\ and F19 are also given. For APOGEE {and GALAH}, we provide similar statistics differences for the whole comparison sample and when only their stars with \CeFe>-0.15~dex are considered. {We used calibrated \gspspec~and APOGEE atmospheric parameters.}}
        \label{fig:Diff}
\end{figure}

In this {\it low-uncertainty sample},
we found 105, 101, {and 44} stars in common with F19, APOGEE, {and GALAH}, all with a \gspspec\ \SNR\ higher than 55. The comparisons between these studies are illustrated in Fig.~\ref{fig:Diff}.
The agreement between this {\it low-uncertainty sample} and F19 is excellent
(no bias and dispersion of 0.15~dex),
whereas the comparison with APOGEE is good with a bias identical to the bias between F19 and APOGEE, together with a larger dispersion of 0.25~dex. 
We note that most of the {GALAH and} APOGEE low Ce abundances seem to be systematically underestimated compared to those from $\gspspec$ and F19.
The agreement with {GALAH and} APOGEE is indeed much better when their lowest Ce abundances are rejected:
keeping only {GALAH and} APOGEE stars with \CeFe>-0.15~dex leads to a mean difference {with respect to \gspspec\ equal to -0.09 and -0.22 dex, respectively. The standard deviation with respect to \gspspec\ becomes equal to 0.18 and 0.18~dex, respectively.}

{The excellent agreement between \gspspec\ and F19 might be explained by the high quality of these spectra: high spectral resolution, and \SNR\ for F19 and high \SNR\ for \gspspec. APOGEE stars in common with \gspspec\ have a slightly lower \SNR.} 

Furthermore, we note that the reported Ce differences cannot be explained by differences in the adopted atmospheric parameters since these {four} studies adopted rather consistent \T\ and \g, as is shown in Fig.~\ref{fig:Diff}. The mean $\Delta$\T\ and $\Delta$\g\ are equal
to {106}~K and {0.22}, respectively, between \gspspec~and APOGEE data {and 30~K and 0.06 between \gspspec~and GALAH data.}
\footnote{A comparison of \gspspec\ \T\ and \g\ with literature values is commented on in \citet{GSPspecDR3}.}

Finally,
the comparison between \gspspec\ and F19 allows us to conclude that both studies are
on the same reference scale (which is not the case for APOGEE), and that no calibration is required to interpret the \gspspec\ \CeFe.

\begin{figure*}[h!]
        \centering
        \resizebox{20.cm}{10.5cm}{\includegraphics{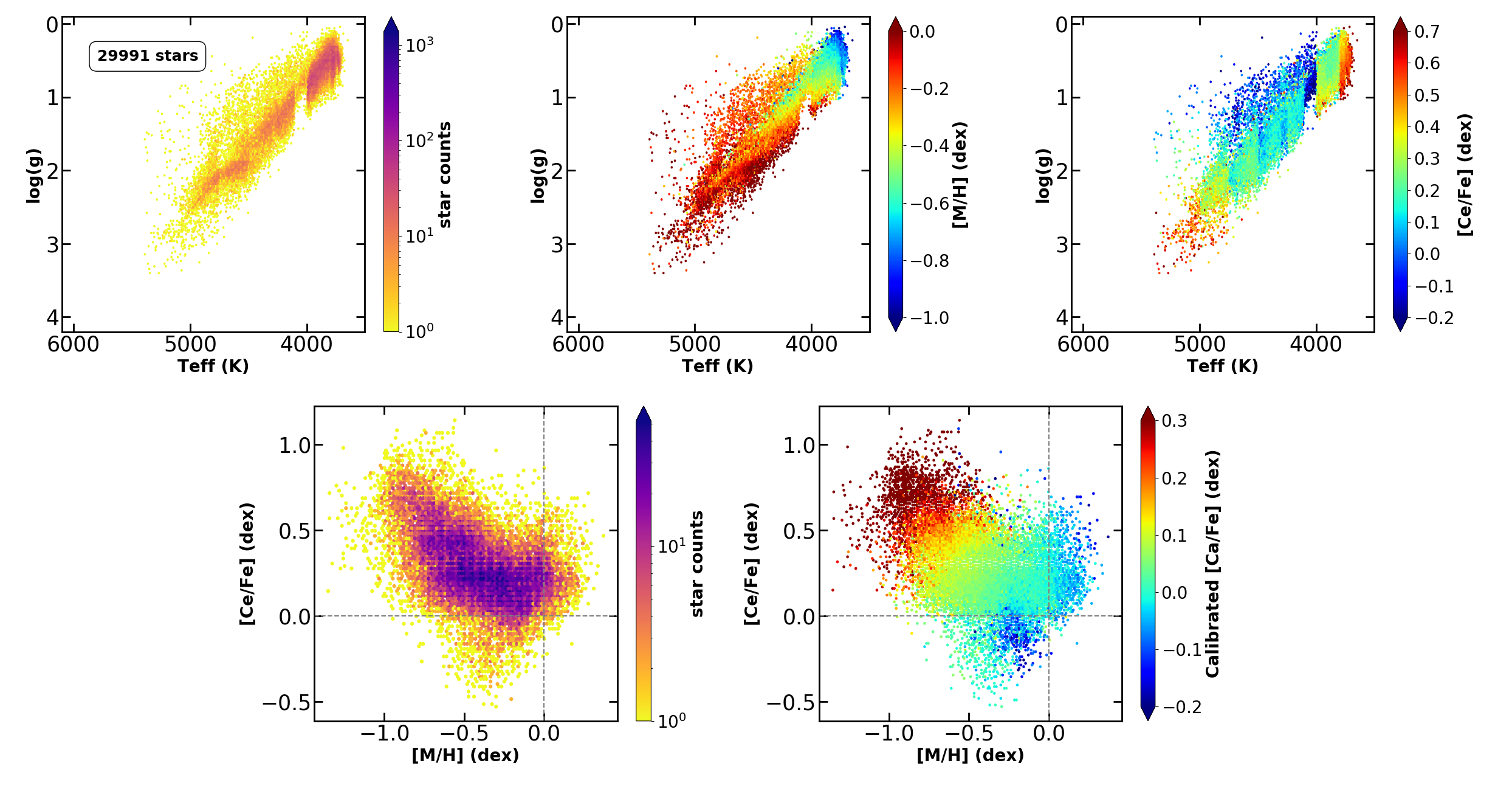}}
        \caption{Upper panels: Kiel diagram of the {\it low-uncertainty sample} stars colour-coded with stellar counts (left panel),  medium metallicity per point (central panel), and medium cerium abundances (right panel). Lower panels: \CeFe\ vs \meta\ distribution colour-coded in stellar counts (left panel) and median calibrated calcium abundances (right panel).}
        \label{fig:Ce-M-Ca}
\end{figure*}

\begin{figure*}[h!]
        \centering
        \resizebox{20.cm}{10.5cm}{\includegraphics{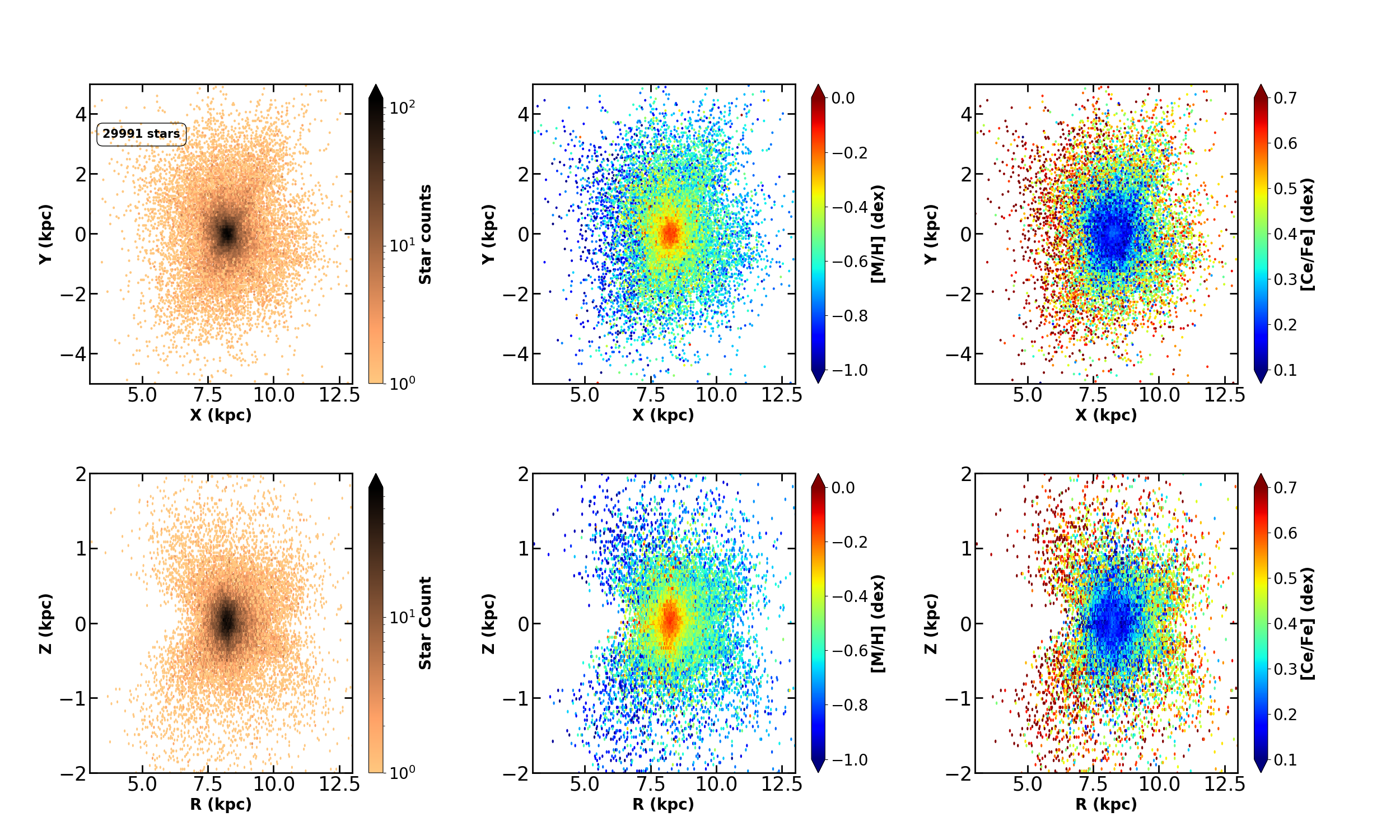}}
        \caption{Galactic distributions of the {\it low-uncertainty sample} stars. The upper panels show the distributions in Cartesian coordinates ($X, Y$), colour-coded from left to right by stellar counts, median metallicities, and median cerium abundances. The bottom panels show the $(R,Z)$ distributions 
        with similar colour-coding as the top panels.}
        \label{fig:Ce-XYRZ}
\end{figure*}

\begin{figure*}[h!]
        \centering
        
        \includegraphics[scale=0.31]{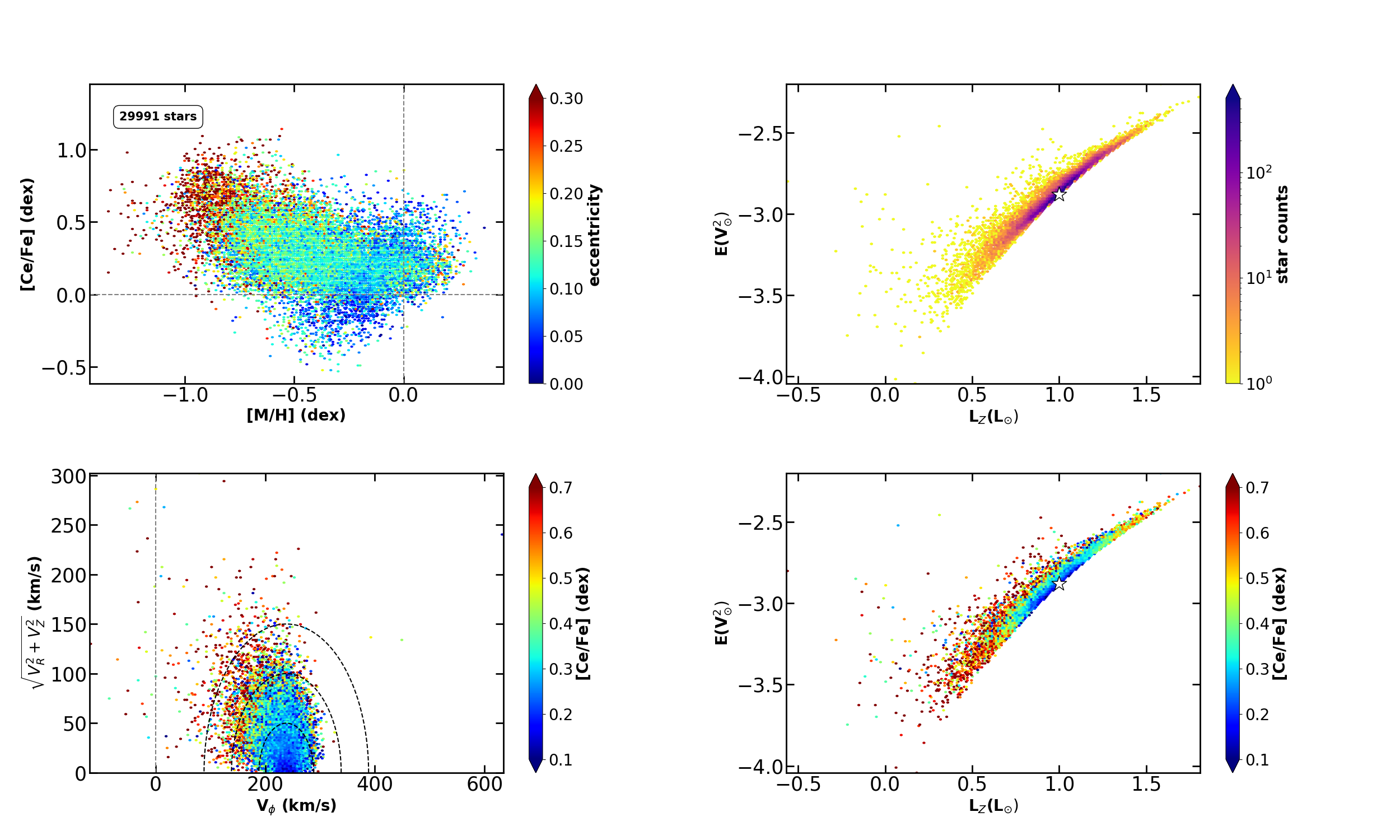}
        \caption{Left upper panel: Distribution of cerium vs metallicity for the {\it low-uncertainty sample} stars, colour-coded with the eccentricity of their orbits. Left lower panel: Toomre diagram colour-coded with \CeFe. The circular dashed lines corresponds to $V_{\rm Tot}$=50, 100 and 150 km/s. Right panels: Energy-angular momentum (E, L$_Z$) plane colour-coded by density (top panel), and medium cerium abundance (lower panel). The white star indicates the solar neighbourhood (L$_Z$ = L$_\odot$, E $\approx$ -2.88 V$_\odot^2$).}
        \label{fig:Ce-eVT}
\end{figure*}

\section{Chemo-kinematics and dynamical properties of selected Gaia cerium abundances}
\label{Sect:ChemoKin}
In this section, we present the distribution of  spatial, kinematical, dynamical, and chemical properties of the {\it low-uncertainty sample} defined in the previous section. The computation of the stellar positions (galactocentric Cartesian coordinates $X$, $Y$, $Z$) as well as the  galactocentric radius ($R$) and cylindrical velocities (V$_R$, V$_Z$ and V$_\phi$) is presented in \citet{PVP_Ale} and is based on the \cite{Coryn21} \Gaia\ EDR3 distances. The orbital parameter calculation (eccentricities, actions, apocenters, pericenters, and $Z_{max}$) is described in \citet{Pedro}. 

\subsection{Chemical distribution}
\label{Sect:Final-chemical}
We first present in Fig.~\ref{fig:Ce-M-Ca} the Kiel diagram (upper panels) of the {\it low-uncertainty sample} stars, colour-coded in stellar counts, metallicity, and cerium abundances (top, central, and bottom panels, respectively). This sample is mainly composed of red giant branch (RGB) and AGB stars with \g ~$<$3.5. This results from the fact that the studied cerium line is more easily detected in cool giant stars. Moreover, the stars with a detected Ce line in the {\it low-uncertainty sample} are increasingly more metal poor the cooler the giants and the lower their gravity. Cerium abundances were indeed derived for any metallicity including solar in stars located on the RGB, whereas only stars with a metallicity lower than $\sim$-0.5~dex are present at the top of the AGB.
This again results from the difficulty of correctly measuring the Ce line in cool star spectra that become more and more crowded by molecular lines as \T\ decreases.
This is also illustrated in the right panel of Fig.~\ref{fig:Ce-M-Ca}, showing that only the coolest AGB stars (\T$\la$3,800~K) with relatively high cerium abundances (\CeFe $\gtrsim$ 0.60~dex) have been measured. Similarly, only Ce-rich stars hotter than \T$\ga$4,800~K are detected. These observational biases are discussed in Sect.~\ref{Sect:Bias}. 
We also remark that there is a lack of stars around \T\ $\sim$ 4000~K. 
This feature is caused by the complexity of the coolest giant spectra. This could lead to some parametrisation
issues that were partly fixed through the  $KMgiantPar$ flag, even if this means rejecting part of these AGB stars
\citep[see][for more details on this flag]{GSPspecDR3}.

The lower panels of Fig.~\ref{fig:Ce-M-Ca} present the trend of cerium abundances with metallicity, colour-coded with stellar counts (left bottom panel) and with calcium abundances (right bottom panel). We adopted the calibrated Ca abundances according to Tab.~4 of \citet{GSPspecDR3} below, using the polynomial of degree 4 as a function of \g\ and a calibration for \g \ values outside the recommended gravity interval set to those computed for the limiting values.
For all the {\it low-uncertainty sample} stars, their $CaUpLim$ and $CaUncer$ flags are equal to zero, and their uncertainties in \CaFe\ are smaller than 0.06~dex.
Over a metallicity range of~1.5 dex, we found a banana-like shape that can be explained by some detection bias. For example, low-Ce metal-poor stars are poorly represented in our sample because the Ce line becomes too weak to be detected in these stellar atmospheres. Similarly, low Ce abundances are more difficult to derive in crowded metal-rich spectra. 

On the other hand, the lower right panel shows that the sample is composed of stars that are more Ca-rich with decreasing metallicity, a consequence of the behaviour of $\alpha$-elements with \meta\ in the Milky Way. In the most metal-poor regime, there are predominantly Ce-rich stars (\CeFe~$\gtrsim$ 0.5 dex) with high calcium abundances (\CaFe~$\gtrsim$ 0.3 dex). 

We finally note an excess of stars with rather low \CeFe\ and \CaFe\ abundances around \meta =-0.5~dex. As explained in \citet{PVP_Ale}, these stars are mostly massive and young (see their Fig.~8) and are located in the spiral arms of the Milky Way. {This is confirmed by the recent work of \citet{Poggio22} (see their Fig. B.1.). Their sample A is a sub-sample containing the majority of the massive star sample defined in Fig. 8 of \citet{PVP_Ale}. These stars present very young ages ($<$ 500 Myr) and are massive.}

There are also some other stars with low \CeFe\ and solar \CaFe\ that probably belong to the disc because they have similar spatial distribution and kinematics as other disc stars.

\subsection{Spatial distribution}
We illustrate in Fig.~\ref{fig:Ce-XYRZ} the spatial distribution of the {\it low-uncertainty sample} stars. Three maps in the ($X, Y$) plane are shown in the top panels. These maps are colour-coded by stellar counts, median metallicity, and median \CeFe\ (from left to right). The bottom panels of Fig.~\ref{fig:Ce-XYRZ} show the same sample stars in the $(R, Z)$ plane.
We first remark that the spatial coverage is quite large (about 7~kpc in $X$, $Y$ and $Z$), even though most of the sample is concentrated in the solar neighbourhood. 
However, it is worth noting that this figure shows (and comparison of it to the Kiel panels of Fig.~2) that the most metal-rich stars have preferentially larger \g\ are found closer to the Sun, while the most metal-poor stars are more likely to be giants and can be seen out to larger distances. This results from some observational biases that are treated in Sect.\ref{Sect:Bias}

In addition to these possible biases, the closest stars, which are rather more metal rich and more Ce poor than the more distant stars, probably belong to the thin disc as they are mainly concentrated within $\pm$0.5~kpc from the Galactic plane. Moreover, their $Z_{\rm max}$ is lower than 0.8 kpc for about 90\% of them. There could thus nevertheless be a small contribution from thick-disc or halo stars in this sample.
On the other hand, stars with higher Ce abundances are found to be more metal poor (see Fig.~\ref{fig:Ce-M-Ca}) and are preferentially located outside the solar neighbourhood and/or at larger distances from the Galactic plane. Part of this population probably does not belong to the thin disc because it is located at
$|$Z$|>$1~kpc. This agrees with their metallicity and Ca content (see the bottom right panel of Fig.~\ref{fig:Ce-M-Ca}). 

\subsection{Chemo-kinematics and chemo-dynamics}
Based on the kinematical and orbital parameters presented in \citet{Pedro}, we show in the upper left panel of Fig.~\ref{fig:Ce-eVT} the Ce abundances with respect to the metallicity, colour-coded with the median eccentricity of their Galactic orbit. 
We remark that the stars with higher Ce abundances orbit on more eccentric orbits. This confirms that these stars probably do not belong to the thin disc. In contrast, more metal-rich stars with a Ce abundance are on almost circular orbits with $Z_{\rm max}$ smaller than $\sim$800~pc, which is typical of thin disc stars.

These trends are confirmed by the Toomre diagram of the {\it low-uncertainty sample} stars (left bottom panel of Fig.~\ref{fig:Ce-eVT}) colour-coded by the median \CeFe. Ce-enriched stars are mostly outside the annulus of 150 km/s, suggesting that they belong to the Galactic halo and/or thick disk. We can also see that a huge majority of these stars (95.4 \%) exhibit disc kinematical properties since their total velocity is always lower than $\sim$100~km/s. This is confirmed by their $Z_{\rm max}$
, which is smaller than 800~pc for $\sim$ 85\% of the {\it low-uncertainty sample}. 
Despite this dominance of the disc population, it is worth noting that a smaller proportion of halo stars, including objects on retrograde orbits, is also present.

Finally, the right panels show the total energy $E$ (rescaled in terms of V$_\sun^2$) with respect to the vertical component of the angular momentum L$_Z$ (fixed as positive in the direction of Galactic rotation), colour-coded in stellar counts (upper panel) and \CeFe\ abundances (lower panel).  These plots again confirm that the large majority of the {\it low-uncertainty sample} stars is located inside the Galactic disc, and  especially close to the Sun (indicated by the white star in the figure). 
In addition, lower angular momentum halo stars can be observed. Some of these stars fall into already identified halo substructures such as the \Gaia-Enceladus-Sausage  \citep[GES; see][]{Helmi18, Belokurov18, Myeong18, Feuillet20, Feuillet21} 
at low $|L_Z|$ and $-2.8V_{\odot}^2 \lesssim E \lesssim -2.0V_{\odot}^2 $. These stars are discussed in Sect.\ref{SectAccret} by extending the analysis to the {\it complete sample}.

\section{Cerium in the Galactic disc}
In this section, we discuss the chemical evolution of cerium in the Galactic disc based on these \gspspec\ data. For this purpose, it has to be taken into account that the {\it low-uncertainty sample} defined in Sect.~\ref{Sect:Sample} could be biased by some selection function effects, for instance, spatial distribution and stellar parameter limitations. In particular, the \gspspec\ cerium line cannot be detected and measured for any combination of stellar atmospheric parameters. Ce abundances are indeed available only for giant stars, as
already shown in Fig.~\ref{fig:Ce-M-Ca}. Moreover, only the brightest AGB stars located far outside the solar neighbourhood can have a derived \CeFe. Similarly, we showed that cerium abundances are measured with difficulty in crowded spectra of metal-rich and/or very cool stars. This could favour the detection of Ce-enriched stars, not always representative of the ISM Galactic content, due to the modification of the atmospheric s-element abundances in evolved low-mass stars caused by the internal production. To take these biases into
account, we defined a new stellar subsample (called {\it high-quality sample}, hereafter) in order to discuss Galactic Ce gradients and trends. The Galactic evolution of this neutron-capture element is then interpreted based on a chemical evolution model. Finally, we explore the \CeFe~abundances in open clusters by tracing Galactic gradients complementary to field stars.

\subsection{High-quality sample of Ce abundances}
\label{Sect:Bias}
To consider the most accurate Ce abundances (low uncertainties and
best stellar parametrisation) and to avoid detection biases towards more Ce-rich stars, we selected only results for \SNR~$\ge$~300 and \CeFe\ uncertainty $\le$~0.10~dex. Then,
since the Ce line is more easily detected in cool stars, we kept only stars with 3,800 $\le$\T$\le$ 4,800~K (as discussed from Fig.~\ref{fig:Ce-M-Ca} and associated text). On one hand, hotter star spectra have a very weak and almost undetectable Ce line ({as already shown in Fig. \ref{fig:CeKiel}}), thus only Ce-rich stars can be measured at these temperatures. On the other hand, the cut at low \T\ rejects the coolest AGB stars of the sample, most of them being metal poor and Ce rich (see the top panels of Fig.~\ref{fig:Ce-M-Ca}). These stars are probably enriched in Ce due to their internal nucleosynthesis and mixing. 
Their properties will be discussed in a future article.

Finally, the {\it high-quality sample} is composed of 7,397 stars mainly located within 1~kpc from the Sun in $X-Y$ coordinates. The left panel of Fig.~\ref{fig:RZ-Toomre} shows their location in the ($R_g$-$Z_{max}$) plane. Only a few of them have |$Z_{max}$| > 0.7~kpc ($\sim$10\% of the sample). Their Galactic velocities are compatible with a membership to the disc, as can be deduced from the Toomre diagram presented in the right panel of Fig.~\ref{fig:RZ-Toomre}, which shows that $\sim$ 85\% of them have a total velocity lower than $\sim$70~km/s and a $Z_{\rm max}$ smaller than 700~pc.

\subsection{\CeFe\ versus \meta\ trends}

\begin{figure}
        \centering
        \includegraphics[scale=0.15]{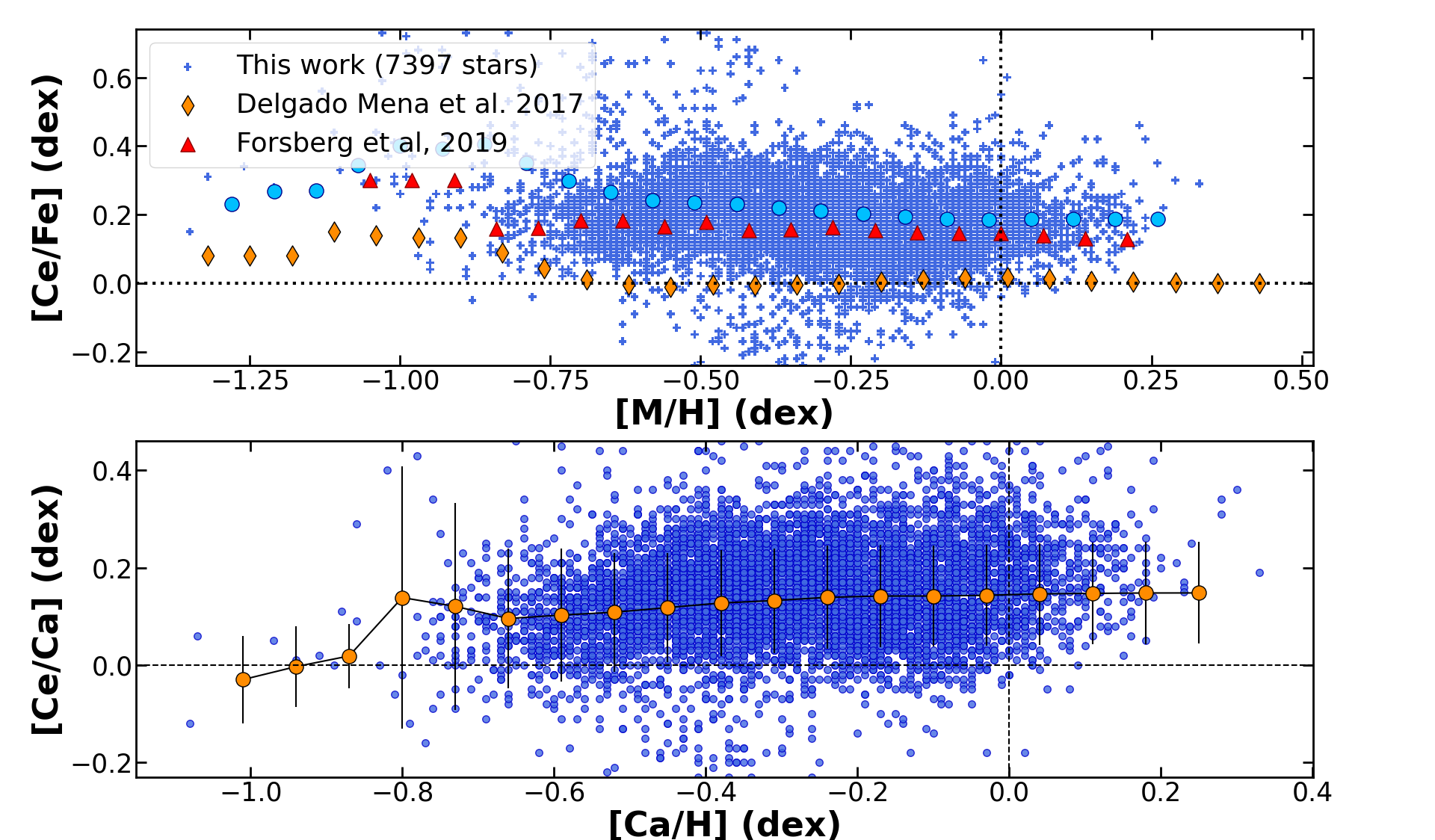}
        \caption{Top panel: Cerium and iron abundances ratio for the {\it high-quality sample} with respect to the metallicity. Red triangles and orange diamonds are mean \CeFe\ ratios for the stars of F19 and \citet{DM17}, respectively (computed per bins of 0.07 dex). Sky blue points the mean of our data per bin of 0.07 dex in [Ca/H]. Bottom panel: [Ce/Ca] vs [M/H]. Orange dots correspond to the mean of the measurements per bin of 0.07~dex, and the error bars correspond to the standard deviation in each bin.}
        \label{fig:Ce-Lit}
\end{figure}

\begin{figure*}
        \centering
       
        \includegraphics[scale=0.32, width=1.0\textwidth, height = 11 cm]{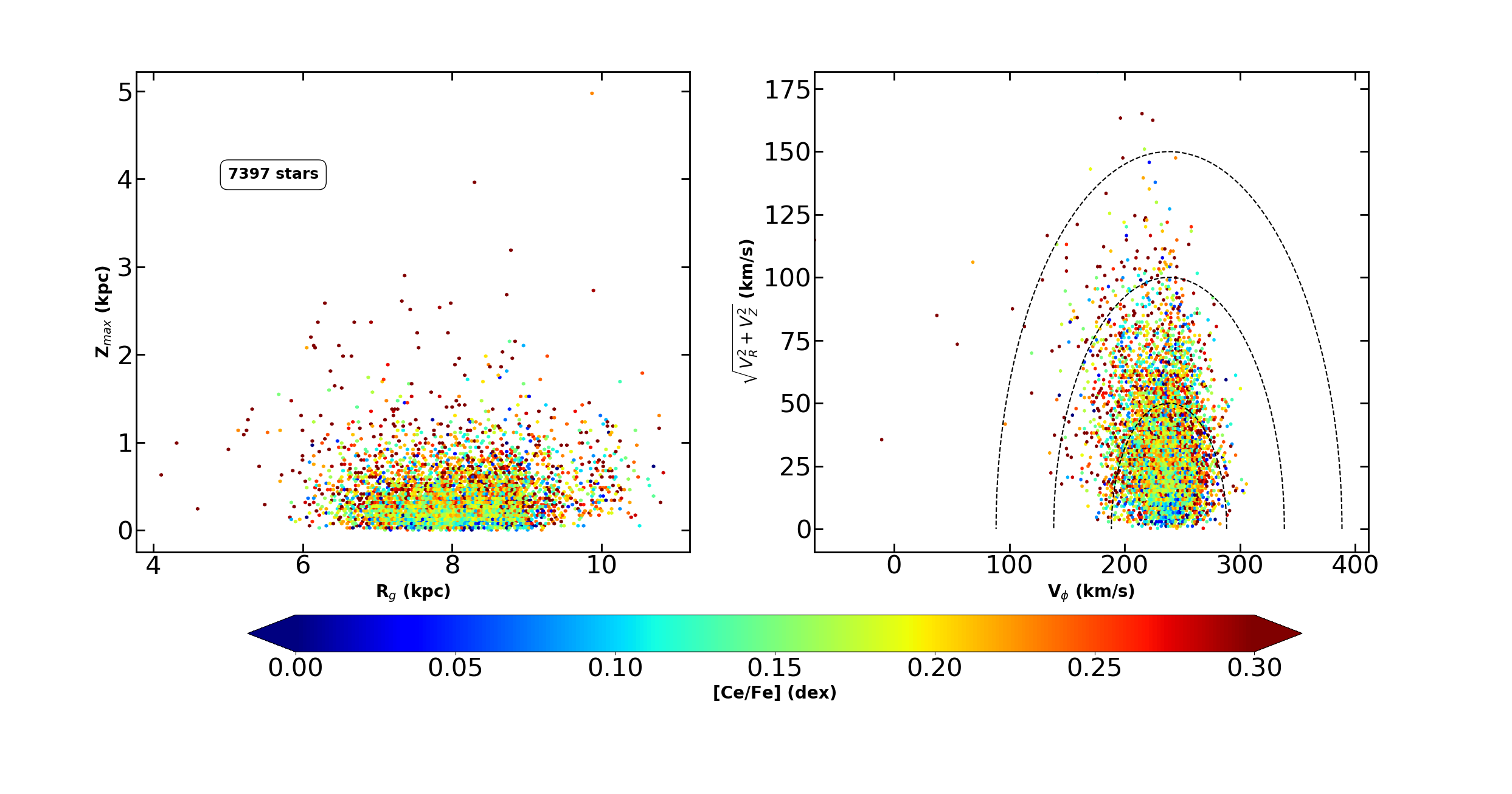}
        \caption{Left panel: Galactic distribution in the ($R_g$-$Z_{max}$) plane of the {\it high-quality sample} stars colour-coded with the median cerium abundances per point. Right panel: Toomre diagram of the same stars.}
        \label{fig:RZ-Toomre}
\end{figure*}

In order to validate this {\it high-quality sample}, we illustrate the \CeFe\ trend with respect to  metallicity in the top panel of Fig.~\ref{fig:Ce-Lit}. We found a rather flat trend at a mean level of \CeFe$\sim$0.2~dex for metallicities varying between $\sim$-0.7 up to $\sim$+0.3~dex. A similar behaviour and mean level of \CeFe\ is reported by F19, based on 277 stars (red triangles in Fig.~\ref{fig:Ce-Lit}, top panel).
This flat trend also agrees with \citet{Reddy06} (178 stars), \citet{BB16} (365 stars), and \citet{DM17} (orange diamonds in Fig.~\ref{fig:Ce-Lit}, top panel, 653 stars. {These stars have \T~$>$ 5300 K and \SNR~$>$ 100, according to their Sect.~4)}, although these authors report a lower \CeFe\ level ($\sim$+0.0~dex), probably resulting from different calibrations and/or reference scales.
Finally, it is worth noting that in the low-metallicity regime {(\meta~$\la$~-0.8 dex)}, the {\it high-quality sample} is probably not statistically representative.

The bottom panel of Fig.~\ref{fig:Ce-Lit} shows the [Ce/Ca] abundance ratio versus [Ca/H]. Orange dots again illustrate the running mean [Ce/Ca] abundance in bins of 0.07~dex in [Ca/H]. Error bars are the associated standard deviation for each bin. For values of [Ca/H] higher than $\sim$-0.7~dex
(low statistics blur the trend at lower metallicities), we found a slightly increasing [Ce/Ca] abundance with increasing [Ca/H] ($\delta$[Ce/Ca]/$\delta$[Ca/H] = 0.087$^{\pm 0.013}$), similar to the trend of the high-Ia  population of \citet{Griffith21} (this population represents their thin-disc low-\MgFe\ distribution).  It is important to note that \citet{Griffith21} used Mg abundances from APOGEE DR16 data, while the $\alpha$-element reference is Ca in our study.\footnote{We adopted Ca instead of Mg as several of the {\it high-quality sample} stars lack \gspspec\ magnesium abundances.} This continuous increase in [Ce/Ca] could be the consequence of the later contribution of AGB stars (main producers of $s$-process elements, such as cerium) in the Galactic chemical evolution history with respect to SN~II (producers of $\alpha$-elements as Ca). Moreover, we point out that we also found a rather flat distribution of the [Ce/Ca] ratio for [Ca/H] > 0.1~dex, whereas \citet{Griffith21} reported a strong decrease. This is due to the different trend of our Ca and their Mg abundances. Their [Mg/Fe] remains constant for positive metallicities, in contrast to the continuous decrease in our [Ca/Fe], as shown in Fig.25 of \citet{GSPspecDR3}. This continuous decrease agrees better with Galactic evolution models that predict a similar decrease in any \alphaFe\ ratios with \meta. We also note that our [Ce/Ca] is systematically higher than that of \citet{Griffith21}, probably because of the different reference scales that were adopted. 

Finally, we emphasize that this Fig.~\ref{fig:Ce-Lit} and our conclusions are not modified when the calibrated metallicities proposed by \cite{PVP_Ale} are adopted. We therefore decide to not calibrate \meta\ in the following.

\subsection{Cerium abundance radial gradients.}
\label{Sect:Gradients}
The radial Ce abundance gradients of the Galactic disc were computed using a Theil-Sen fit of the {\it high-quality sample} trends with respect to i) the radial distance to the Galactic centre $R$, and ii) the guiding radius ($R_g$, approximated by the mean of the orbital apocentre and pericentre distances). {Errors were computed by adopting a confidence level of 0.95}.

First of all, we examined the \CeFe\ radial gradients, and flat gradients with respect to $R$ or $R_g$ were found. The corresponding  fits are $\delta$[Ce/Fe]/$\delta$R = -0.001$^{\pm 0.004}$ dex.kpc$^{-1}$ for the Galactic radius, and  
$\delta$[Ce/Fe]/$\delta R_g$ = -0.001$^{\pm 0.005}$ dex.kpc$^{-1}$ for the Galactic guiding radius.

We therefore find that the ISM \CeFe\ content is constant for $R_g$within 7.5 and 9.5~kpc from the Galactic centre. This flat gradient within the Galactic disc has a smaller slope (although it almost agrees within the error bars)  with respect to the gradient in $R_g$ reported by \citet{Taut21}, assuming that their {\it mean galactocentric distances} are equivalent to our $R_g$. Considering only their thin-disc stars, they indeed found a radial gradient of 0.015${\pm 0.007}$ dex.kpc$^{-1}$ over a similar range in R$_g$. 
We found 32 stars in common between the {\it high-quality sample} and \citet{Taut21}, with a mean difference in \CeFe\ and a standard deviation of -0.25 and 0.16~dex, respectively. The two studies therefore agree well, although they are not on the same reference scale. However, this different scale do not affect the cerium gradient determination.
As a consequence, the difference between the two derived gradients might be explained by the smaller number statistics of the \citet{Taut21} study, which relied on only 424 stars  (i.e. less than 6\% of the \gspspec\ sample). As a consequence, selection function biases might be more important in this last study. 

Secondly, the [Ce/H] radial gradients were derived and were found to be marginally negative: $\delta$[Ce/H]/$\delta R = -0.028^{\pm 0.017}$ dex.kpc$^{-1}$ and $\delta$[Ce/H]/$\delta R_g = -0.051^{\pm 0.007}$ dex.kpc$^{-1}$. They are fully consistent with the [Ce/H] and [La/H]\footnote{La is a second-peak $s$-process element that shares a similar production history as Ce \citep[see e.g.][]{Nikos18}.} horizontal gradients ($\delta$[Ce/H]/$\delta R = -0.024^{\pm 0.003}$ dex.kpc$^{-1}$ and $\delta$[La/H]/$\delta R = -0.020^{\pm 0.003}$ dex.kpc$^{-1}$) derived from Cepheids by \citet{DaSilva16} over a wider range of galactocentric distances (4-18 kpc). Our result also agrees with the negative radial metallicity gradients of the disc reported for Gaia data by \citep[see][Sect.~4]{PVP_Ale}.

\subsection{Cerium abundance vertical gradients}
On one hand, the \CeFe\ vertical gradient was derived with respect to the absolute distance to the Galactic plane |Z|.
A positive trend  was found with  $\delta$[Ce/Fe]/$\delta Z=0.122^{\pm 0.016}$dex.kpc$^{-1}$. 
On the other hand, the gradient with respect to the maximum orbital distance to the plane, $Z_{max}$, provides a rather similar value: $\delta$[Ce/Fe]/$\delta Z_{max}$ = 0.086 $^{\pm 0.011}$ dex.kpc$^{-1}$. These two gradients are not affected when stars located at distances larger than $\sim$600~pc from the plane are rejected.
An opposite trend was found by  \citet{Taut21} ($-0.034{\pm 0.027}$~dex.kpc$^{-1}$) for their thin-disc gradient,
probably due to the selection function biases we discussed above. 
Nevertheless, we note that these authors found a positive gradient for La in the thin disc ($\delta$[La/Fe]/$\delta R_g = 0.030^{\pm 0.025}$ dex.kpc$^{-1}$) in better agreement with our vertical gradient and surprisingly in contrast with their Ce gradient. 

Finally, we found decreasing [Ce/H] vertical gradients:  $\delta$[Ce/H]/$\delta Z = -0.453{\pm 0.035}$ dex.kpc$^{-1}$ and $\delta$[Ce/H]/$\delta Z_{max} = -0.297{\pm 0.021}$ dex.kpc$^{-1}$. These are related to the  vertical metallicity gradient that we derived: $\delta$[M/H]/$\delta Z = -0.614{\pm 0.032}$ dex.kpc$^{-1}$.

\subsection{Comparison with Galactic evolution models}

 \citet{spitoni2022} presented a new chemical evolution model designed to reproduce  \gspspec\  [X/Fe] versus [M/H] abundance ratios, where $X$ stands for several $\alpha$-elements in the solar vicinity.
This model is an extension of  recent two-infall models  \citep{spitoni2020,spitoni2021}  designed to reproduce APOKASC  and APOGEE data assuming that  high- and low-$\alpha$ sequence stars are formed by means of two independent episodes  of gas infall.
However, \cite{PVP_Ale} clearly showed a young  chemical impoverishment in metallicity and with low [$\alpha$/Fe] values.   In the new model proposed by \citet{spitoni2022}, this population is  well traced when the low-$\alpha$ population is generated by two sequential infall episodes.  It is worth mentioning that this model is also able to reproduce the star formation history as constrained by previous \Gaia\ releases \citep{bernard2017,lara2020}. In conclusion, an original {\it three-infall} chemical evolution model for the disc components has been proposed. {This {\it three-infall} model is also motivated by the recent work of  \citet{DeCia21}, who highlighted the recent infalling gas of pristine chemical composition in the interstellar medium. \citet{vincenzo2020} showed important signatures of recent metal-poor gas accretion from Milky Way-like simulations in the cosmological framework  ( 0-2 Gyr ago).}
We refer to  Table 2  of \citet{spitoni2022} for the values of  the adopted  model parameters.  
\begin{figure}
\begin{centering}
\includegraphics[scale=0.4]{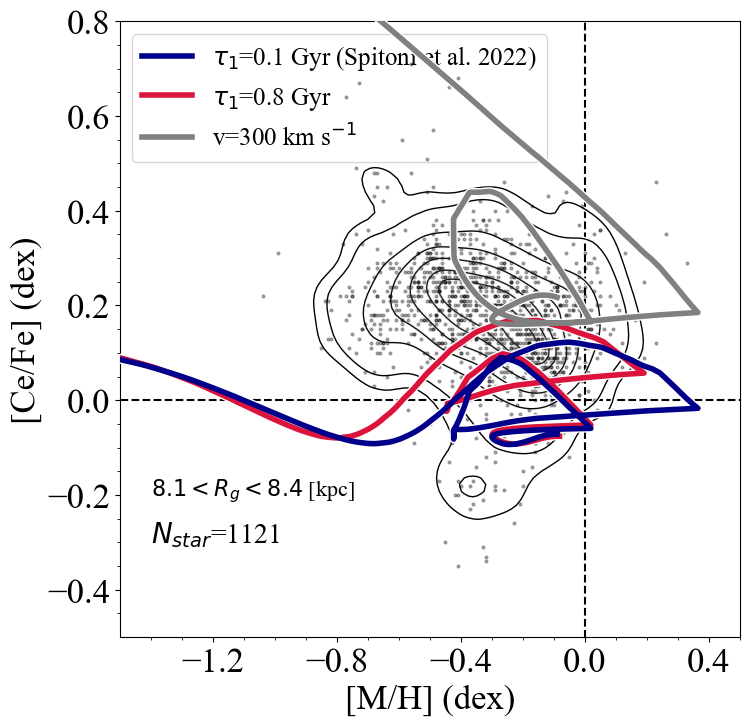}
      \caption{Model predictions for the chemical evolution of Ce in the [Ce/Fe] vs [M/H] plane ({\it high-quality sample}). The blue line stands for the model assuming the  same parameter as in \citet[][see their Table 2]{spitoni2022}. In this case, the timescale of the gas accretion for the high-$\alpha$ sequence is $\tau_1=0.1$ Gyr. The red line shows the case with $\tau_1=0.8$ Gyr. 
      The grey line represents the model where the yields of  \citet{limongi2018}  for rotating massive stars  assuming that all stars rotate with an  initial velocity  of 300 km s$^{-1}$ have been considered. \gspspec\ stars  with guiding radii $R_g$ between 8.1 and 8.4 kpc  are indicated with grey points.
      The contour lines enclose fractions of 0.95,
0.90, 0.75, 0.60, 0.45, 0.30, 0.20, and 0.05 of the total number of observed stars.}
        \label{fig:model}
        \end{centering}
\end{figure}

\begin{figure*}
\begin{centering}
\includegraphics[scale=0.14, width=1.0\textwidth]{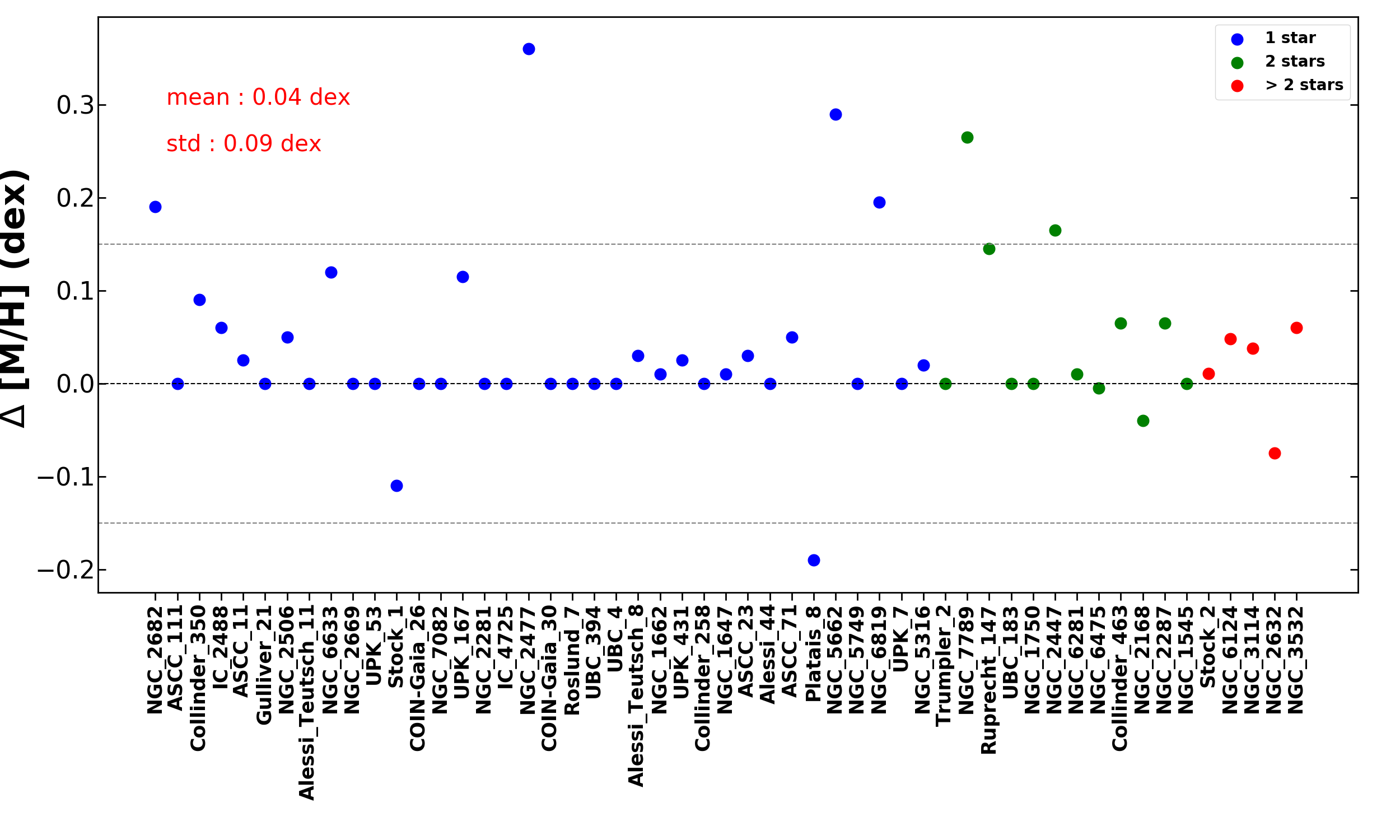} 
      \caption{Difference between literature and \gspspec\ open cluster metallicities (computed as the mean \meta\ of their member) for OC with at least one cerium abundance. Blue, green, and red points indicate the number of stars belonging to each OC (one, two, or more than two members, respectively). The two horizontal lines at $\pm$0.15~dex indicate the OC with good metallicities that are kept for further analysis. }
        \label{fig:OC-Mcheck}
        \end{centering}
\end{figure*}

\begin{figure*}
\begin{centering}
\includegraphics[scale=0.14, width=1.0\textwidth]{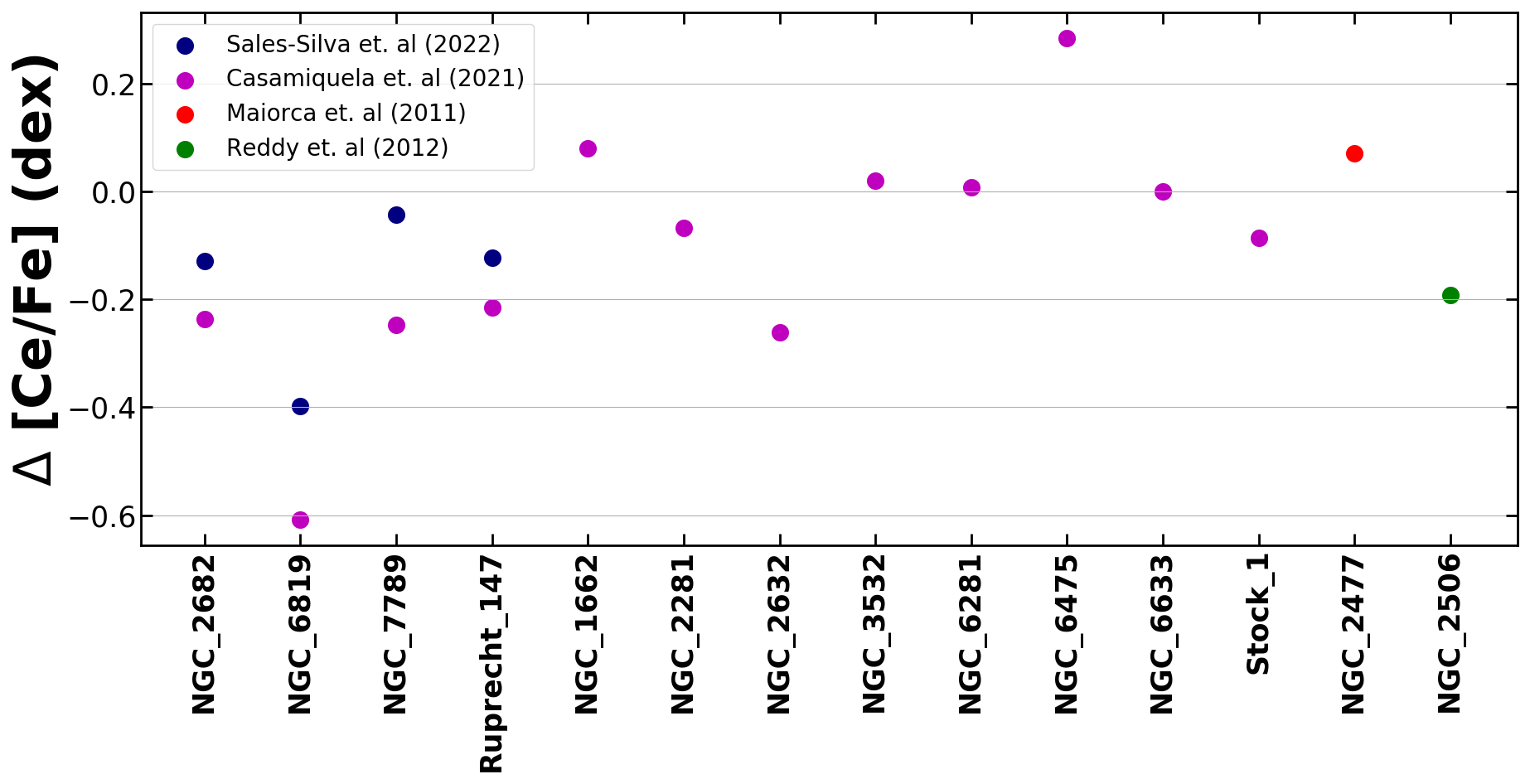} 
      \caption{Difference between literature (blue circles are OC from \citet{SalesSilva22}, purple circles from \citet{Casamiquela21}, red circles from \citet{Maiorca11}, and green circles from \citet{Reddy12}) and \gspspec\ open cluster \CeM~(computed as the mean \CeFe\ of their member) for OC with at least one cerium abundance. }
      
        \label{fig:OC-MCecheck}
        \end{centering}
\end{figure*}

In this section,  we show  predictions of  this  {\it three-infall} model for the chemical evolution of Ce.  
We recall that Ce is formed from both the $s$- and $r$-process channels \citep{Arlandini99}:

\begin{itemize}
    \item 
The most part of the $s$-process in Ce is synthesised in   low-mass AGB stars in the mass range 1.3-3 M$\odot$ , and the corresponding yields are taken  from  \citet{cristallo2009,cristallo2011}

\item  The contribution of the   $s$-process  from rotating massive stars was also taken into account.
First implemented by \citet{cescutti2013,cescutti2014,cescutti2015} considering the nucleosynthesis prescriptions  of \citet{frisck2012}, 
we included  the  yields of \citet{frisch2016},  as indicated in Table 3 of \citet{rizzuti2019}. We also tested the yields produced by rotating massive stars as proposed
by \citep{limongi2018}.
\item The Ce yields produced by the $r$-process
have been computed scaling the Eu yields according
to the abundance ratios observed in $r$-process-rich stars \citep{sneden2008}.
For Eu nucleosynthesis,
 we included the production of Eu from neutron star mergers (NSM). 
 Following the prescriptions of \citet{matteucci2014} and \citet{cescutti2015},  the  value of the NSM yield is  $2\cdot10^{-6}$ M$_{\odot}$  and  the time delay due to the coalescence of the two neutron stars is  equal to 1 Myr.   We refer to Section 3.2.2 of \citet{grisoni2020} for further details.
\end{itemize}

In Fig. \ref{fig:model} we compare our model predictions for [Ce/Fe] versus [M/H]  abundance ratio in the solar vicinity with the {\it high-quality sample} stars defined above. We also considered  only stars with guiding radii $R_g \in [8.1, 8.4]$ kpc, consistent with the \citet{spitoni2022} stellar samples.

First, we recall that \citet{grisoni2020} followed  the evolution  of the Galactic thick and thin discs with  a parallel approach  \citep{grisoni2017} by means of two distinct infall episodes evolving separately (i.e. two distinct chemical evolution tracks in the [Ce/Fe] versus [Fe/H] space). Our model predictions agree with the  findings reported in Fig. 2  of  \citet{grisoni2020} for the high-$\alpha$ sequence, although less Ce-rich stars are predicted when compared to the \gspspec\ observations (see discussion below). 
On the other hand, for low-$\alpha$ stars, the chemical dilution from gas infall episodes (which create two loop features in the [Ce/Fe] versus [M/H] ratio plane) is absent in the \citet{grisoni2020} prediction.

 The most recent dilution event, which started $\sim 2.7$~Gyr ago,  has the main effect of impoverishing the metallicity of the younger stellar populations \citep[see][]{PVP_Ale} and also allows us to predict the young population at subsolar [Ce/Fe] and [M/H] values seen in Fig.~\ref{fig:model}.

In addition, we tested the effects on the model of different values for the timescales of gas accretion in the high-$\alpha$ sequence assuming 0.1 Gyr (as in \citealt{spitoni2022}) and 0.8 Gyr.
We note that a longer timescale helps  to better reproduce the data by predicting higher [Ce/Fe] values, as observed. Different nucleosynthesis prescriptions to improve the agreement between the {\it three-infall} model and the observed \CeFe\ even more,\ especially some stars with higher [Ce/Fe] values, will be considered in a  future work. But we can already conclude that the {\it three-infall} chemical evolution model well reproduces the observed \CeFe\ abundances if a longer results  time-scale for the last gas accretion is considered.

Finally, we stress that our results also agree with the  model predictions  of \citet{Nikos18} , who considered the  yields of massive stars of \citet{limongi2018}  weighted by a metallicity-dependent function of the  rotational velocities. 
In Fig.~\ref{fig:model} we also show   the results for an extreme case. We assumed  the same parameter as in \citet{spitoni2022}, but considering  the contribution of rotating massive stars of \citet{limongi2018},   where all stars rotate with the highest initial velocity of 300  km s$^{-1}$. We are aware that this choice  for all stars formed  at all metallicities  is not physically  motivated. However, Fig.~\ref{fig:model} shows that  a larger   contribution of the highest  velocity stars    could improve the agreement with the data presented in this work. Nevertheless, as shown in \citet{rizzuti2019}, this extreme nucleosynthesis prescription overproduces the ratio of [Ba/Fe] and [Sr/Fe]  abundance ratios.

\begin{figure*}
\begin{centering}
\includegraphics[scale=0.32]{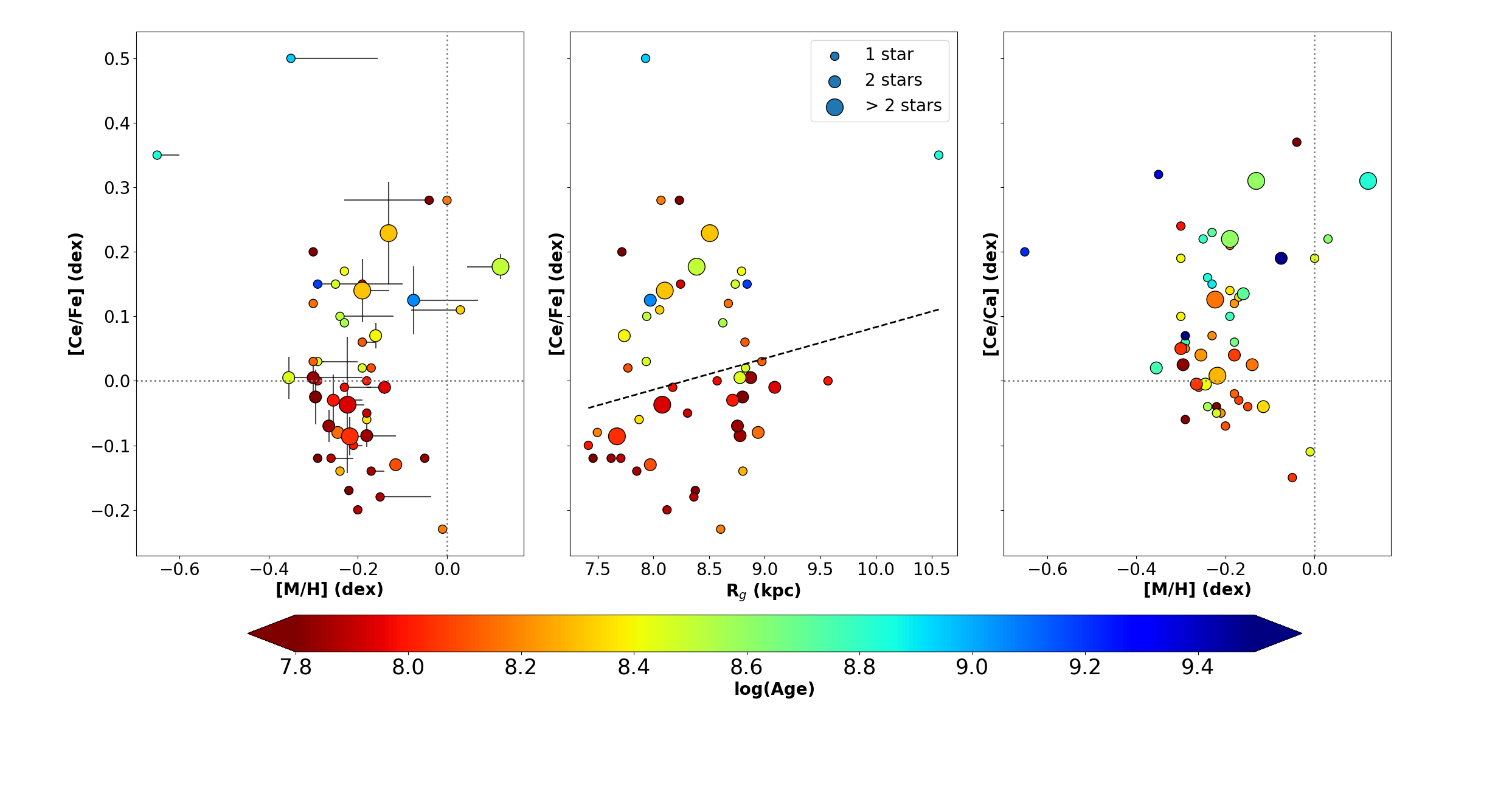}
      \caption{Left panel: \CeFe\ vs \meta~colour-coded with the OC ages. Each \CeFe\ and \meta\ value is the mean of all the OC members. Vertical error bars are the mean dispersion in cerium abundances (star-to-star scatter), and the horizontal lines link the \gspspec\ \meta\ to the reference values. Central panel: [Ce/Fe] abundances with respect to the guiding radius colour-coded with age. The dashed line illustrates the derived radial gradient. {Left panel : [Ce/Ca] ratio with respect to the metallicity colour-coded with the age.} }
      
        \label{fig:OC}
        \end{centering}
\end{figure*}

\subsection{Cerium in open clusters}

To trace the Ce content in the Galactic disc, we also searched for stars belonging to Galactic open clusters (OC).  Since rather few OC members were found within the {\it high-quality sample},
we adopted the {\it low-uncertainty sample}. To select the OC members, we proceeded as in \citet{PVP_Ale} by adopting OC properties and, in particular, ages, from \citet{CG20}, \citet{CG21} and \citet{T21} {These parameters were derived from stars with a membership probability $\ge$ 0.7}. The adopted Galactocentric distances for the clusters are those of \citet{PVP_Ale}. We found 82 stars with \CeFe\ estimates belonging to 53 different OCs. Thirty-six of these OCs have only one member in our sample, 12 OCs have two members, and 5 OCs have three or more members. {Table \ref{Tab:OC} contains the mean [M/H], [Ce/Fe], [Ca/Fe], and number of stars for our 53 open clusters.} 

First, the mean metallicity and associated standard deviation for each cluster were derived. Fig.~\ref{fig:OC-Mcheck} compares the \gspspec~mean metallicities for each of them with that of the above-mentioned catalogues. Since these mean \meta\ were estimated from the stars with a Ce abundance, this OC sample is biased by the Ce line detection: for instance, there is a lack of metal-rich clusters.
In any case, an excellent agreement is found, with a mean metallicity difference of 0.04~dex and a standard deviation of  0.09~dex, which again confirms the high quality of the \gspspec\ chemical analysis. The metallicity of only three open clusters differs by more than 0.2~dex with respect to the reference value, two of which have only one member (their metallicity difference is around 0.3~dex).
In the following, we have kept only the 46 OC with a \meta\ difference with respect to the literature within $\pm$0.15~dex. These good metallicities should be associated with our best derived \CeFe. We also note that the global accuracy in \meta\ shown in Fig.~\ref{fig:OC-Mcheck} was very slightly improved when we applied the calibration in metallicity proposed in Tab.~3 of \citet{GSPspecDR3}, but the precision remained the same.

We therefore decided to calibrate these metallicities no longer.
{Fig.~\ref{fig:OC-MCecheck} compares the \gspspec~\CeM~values with OC literature measurements: \citet{Maiorca11} (red circles), \citet{Reddy12} (green circle), \citet{Casamiquela21} (purples circles), and \citet{SalesSilva22} (blue circles). The differences between these literature studies and \gspspec~cerium abundances are $\Delta$[Ce/Fe]~= 0.07 $\pm$ 0.00, -0.19 $\pm$ 0.0, -0.11 $\pm$ 0.21, and -0.17$\pm$0.13, respectively. We remark that we found a relatively good agreement for our cerium abundances, even though the reference level of \citet{Reddy12} and \citet{SalesSilva22} seems to be different than ours. This last work is indeed on the same scale as the APOGEE DR16 data (they found a mean difference between their [Ce/H] values and that of APOGEE DR16 of 0.05$\pm$0.16). We note that the reference scale of \citet{Maiorca11} is close to ours, but the difference of 0.07 dex between \gspspec~data and that of \citet{Maiorca11} could be explained by the different solar abundances they adopted. They found super-solar abundances of Ce (and elements mainly produced by s- process, e.g. Y, Zr, and La) for their younger OC (with age $<$ 1.5 Gyr). }

We illustrate the behaviour of these OC mean \CeFe\ abundances in Fig.~\ref{fig:OC}. The left panel shows \CeFe~versus~\meta~colour-coded with the cluster ages. We first remark that older OCs appear to be more enriched in Ce 
than younger OCs,in contrast to what was found in \citet{SalesSilva22}. 
This might be caused by the \gspspec\ biases that are induced when the Ce lines were analysed. This analysis indeed favours the selection of Ce-enriched cool stars in the {\it low-uncertainty sample},
as we showed in Fig.~\ref{fig:Ce-M-Ca}. 
{When only stars from the {\it high-quality sample} are selected (20 stars belonging to 14 OCs), the relation we found between \CeFe~and age is unaffected. Removing AGB stars or stars whose $vbroad$ $<$ 9 km/s (43 stars) does not affect the relation we found either. Nevertheless, young stars may be affected by chromospheric activity \citep{Spina20}. We found no sign of chromospheric activity for these stars according to Gaia DR3 data ($activityindex\_espcs$ in $gaiadr3.astrophysical\_parameters$), however.}

The {central} panel presents the OC mean [Ce/Fe] abundances with respect to their guiding radius colour-coded with age. Over a guiding radius varying between $\sim$ 7.2 kpc and 10.5 kpc, we found a radial gradient with a very small slope: $\delta$[Ce/Fe]/$\delta R_g = 0.05^{\pm 0.09}$ dex.kpc$^{-1}$. We highlight that this value agrees (within the error bar) with the flat gradient reported in Sect.~\ref{Sect:Gradients} from the analysis of field disc stars. 
We note that removing the most distant cluster does not change the OC gradient significantly ($\delta$[Ce/Fe]/$\delta R_g = 0.00^{\pm 0.07}$ dex.kpc$^{-1}$). As a comparison, \citet{SalesSilva22} reported {an increasing gradient ($\delta$[Ce/Fe]/$\delta R = 0.014^{\pm 0.007}$ dex.kpc$^{-1}$) over a wider range of $R$ ($\sim$6 -- 15~kpc),} which is compatible within the error bars with our gradient. {Finally, our OC radial gradient considering [Ce/H] ($\delta$[Ce/H]/$\delta R_g = - 0.01^{\pm 0.15}$ dex.kpc$^{-1}$) is compatible within the error bars with that of \citet{SalesSilva22}, $\delta$[Ce/H]/$\delta R_g = - 0.070^{\pm 0.007}$ dex.kpc$^{-1}$.}

{Finally, the right panel shows the [Ce/Ca] ratio as a function of metallicity colour-coded with age. No clear trend between [Ce/Ca] versus [M/H] is seen because of the large scatter. 
The youngest OC seem to present lower [Ce/Ca] values, which contradicts with what found in \citet{SalesSilva22}, for instance. To conclude, a further investigation on the biases of our young open clusters could be useful to understand the behaviour we obtained.}

\section{Cerium in the Galactic halo}
As already mentioned in Sect.~\ref{Sect:ChemoKin}, some \gspspec\ stars with cerium abundances belong to the Galactic halo. We explore the properties of some of them below.

\subsection{Cerium in accreted stars}
\label{SectAccret}

\begin{table*}[t]
        \centering
        
        \begin{tabular}{lcccccc}
                \hline
                \hline
                 $Gaia$ DR3 Id & \SNR & \T~(K) & \g & \meta~(dex) & [Ca/Fe]~(dex) & \CeFe~(dex)\\
                \hline
                {\bf Thamnos} & & & & {\bf -1.26 $\pm$ 0.13} &{\bf  0.26 $\pm$ 0.01} & {\bf 0.59 $\pm$ 0.03}\\
                1294315577499064576 & 657    & 4309 & 1.09 & -1.13 $\pm$ 0.01 & 0.27 $\pm$ 0.01 & 0.56 $\pm$ 0.08 \\
                6423592399737133184 & 102    & 4180 & 0.53  & -1.39 $\pm$ 0.09 & 0.26 $\pm$ 0.02  & 0.62 $\pm$ 0.22 \\
                \hline
                {\bf Helmi Stream} & & & &{\bf -1.18 $\pm$ 0.27} &{\bf 0.25 $\pm$ 0.04} & {\bf 0.32 $\pm$ 0.05} \\
                816615227344979328 & 174 & 3916 & 0.68 & -0.91 $\pm$ 0.02 & 0.29 $\pm$ 0.02 & 0.37 $\pm$ 0.17 \\
                1275876252107941888 & 410 & 4391 & 0.70 & -1.45 $\pm$ 0.01 & 0.21 $\pm$ 0.01 & 0.27 $\pm$ 0.15 \\
                \hline
                {\bf \Gaia-Sausage-Enceladus} & & & & {\bf -1.16 $\pm$ 0.12} & {\bf 0.28 $\pm$ 0.08} & {\bf 0.53 $\pm$ 0.13} \\
                4454379718774068736 & 221 & 4432 & 0.68 & -1.35 $\pm$ 0.01 & 0.23 $\pm$ 0.02 & 0.42 $\pm$ 0.18 \\
                4231500087527853696 & 214 & 4314 & 0.80 & -1.25 $\pm$  0.01 & 0.26 $\pm$ 0.01 & 0.42 $\pm$ 0.14 \\
                810961091879119616 &  93 &4319 & 0.94 & -1.01 $\pm$ 0.04 & 0.21 $\pm$ 0.03 & 0.77 $\pm$ 0.21 \\
                2744053785077163264 & 197 & 4141 & 0.76 & -1.01 $\pm$ 0.15 & 0.06 $\pm$ 0.02 & 0.41 $\pm$ 0.19\\
                3232875420468258432 & 97 & 4250 & 1.50 & -1.11 $\pm$ 0.07 & 0.24 $\pm$0.03 & 0.54 $\pm$ 0.17 \\
                921352299825726208  & 89 & 4126 & 0.66 & -1.20 $\pm$ 0.04 & 0.30 $\pm$ 0.02 & 0.44 $\pm$ 0.21 \\
                614044052605639936  & 195 & 4263 & 0.79 & -1.23 $\pm$ 0.03 & 0.29 $\pm$ 0.02 & 0.68 $\pm$ 0.19 \\
                \hline
        \end{tabular}
                \caption{\label{Tab:Accreted} \SNR, \T, \g, \meta, [Ca/Fe], and \CeFe\ (and their associated uncertainties) for the 11  accreted stars. For these three accreted systems, we also report the mean and standard deviation of their chemical abundances in boldface.}
\end{table*}

Gaia stellar orbits and kinematics have unveiled the considerable proportion of merger debris in the halo \citep[e.g.][and references therein]{Helmi18}, now mixed up with in situ formed objects.
As already mentioned above, a small fraction of the {\it low-uncertainty sample} stars has the chemo-kinematical and dynamical characteristics of halo stars. \cite{PVP_Ale} have explored the Gaia DR3 chemical diagnostics of accretion by analysing the metallicity and \alphaFe\ characteristics of stars in several  overdensities in the (E-L$_Z$) diagram.
To complement this first study, we explored the cerium content of these external systems. 

In order to search for accreted stars with derived cerium abundances and to increase the statistics, we adopted the {\it complete sample} and rejected all stars for which the $KMgiantPar$ flag was equal to unity and $gof$~>~-3.80 to avoid any parametrisation issue.

Then, after cross-matching with the sample of stars in halo dynamical overdensities presented in \citet{PVP_Ale}, we found a total of 17 candidate stars with \gspspec\ Ce abundances, two, six, and nine of which lie within the Thamnos  \citep{Koppelman19, Helmi20}, the Helmi Stream \citep{Helmi99}, and the \Gaia-Enceladus-Sausage \citep[GES,][]{Helmi18, Belokurov18, Myeong18, Feuillet20, Feuillet21} substructures, respectively. 
Only one of them is found within the {\it low-uncertainty sample} ($Gaia$ DR3 1294315577499064576 in Thamnos), the others have larger \CeFe\ uncertainties, as expected for these fainter objects. 
We also verified that none of these candidate stars is affected by the observational biases discussed above, which might favour the detection of Ce-enriched stars.

Two of these 17 candidates from the Helmi Stream have already published chemical abundances from the literature, and their accreted nature has already been reported. \citet{Sheffield12} provided very similar atmospheric parameters for   $Gaia$ DR3 816615227344979328 to the \gspspec\ ones with differences in \T, \g, and \meta\ of 16K, 0.08, and 0.17~dex, respectively. They confirmed its accreted nature through radial velocities combined with chemical diagnostics (this star has a lower [Ti/Fe] abundance ratio than disc stars). We also have an excellent agreement for the atmospheric parameters of  $Gaia$ DR3 1275876252107941888: the highest $\Delta$\T, \g, and \meta \ is 100~K, 0.3, and 0.10~dex, respectively, with respect to \citet{Burris00, Ishigaki13, Mishenina01}. Our cerium abundance (\CeFe=0.27 $\pm$ 0.15~dex) is also fully compatible (within the error bars) with that of \citet{Mishenina01} (\CeFe=0.16 dex) and with other already published  $s$-elements abundances such as barium (Ba, Z = 56) and lanthanum (La, Z = 57) reported by \citet{Burris00} ([Ba/Fe]=0.08, [La/Fe] = 0.15 dex) and \citet{Ishigaki13} ([La/Fe] = 0.25). 
Moreover, \citet{Gull21} recently identified this star as belonging to the Helmi Stream and classified it as being moderately $r$-process enhanced.

\begin{figure}
        \centering
        \includegraphics[scale = 0.14]{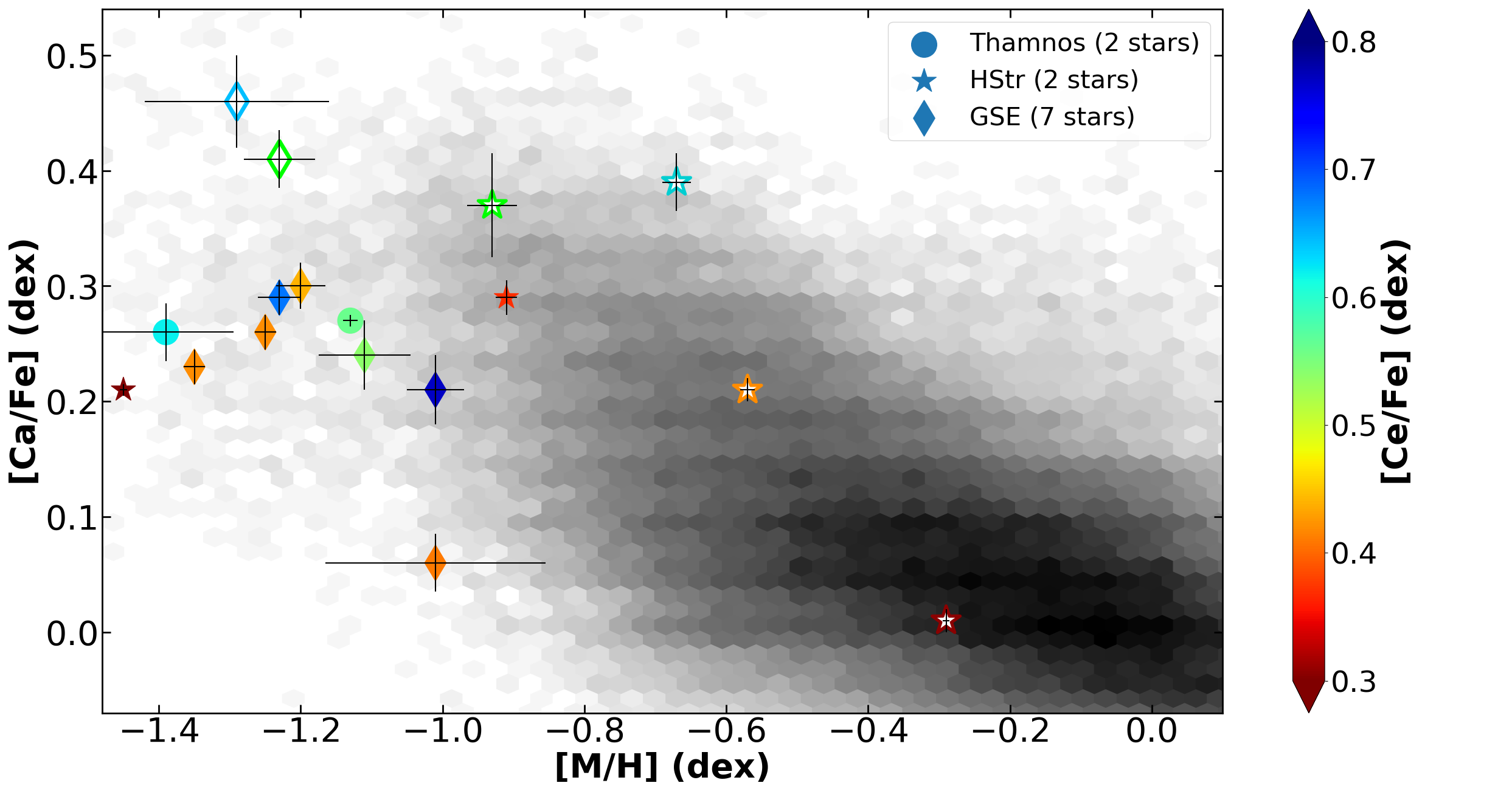}
        \caption{\CaFe\ vs \meta~colour-coded with \CeFe\ abundances for the identified candidate accreted stars. Circles, stars, and diamonds represent stars belonging to Thamnos, the Helmi Stream, and GSE, respectively. The filled symbols refer to the stars that were selected as good member candidates because of their lower \CaFe\ , and the empty symbols are the rejected candidates. The density plot in the background are stars from the solar neighbourhood (see text for more details).}
        \label{fig:Accreted}
\end{figure}

To chemically confirm the accreted nature of the other 15 stars in our sample, we used the Gaia \CaFe\ diagnostic, as already performed in \cite{PVP_Ale} using the \alphaFe\ ratio. Fig.~\ref{fig:Accreted} presents the 17 candidate stars in the (\CaFe\ -- \meta) plane, colour-coded with \CeFe\ abundances. For comparison purposes, the background density plot  illustrates  a selection of high-quality calcium abundances in the solar cylinder, as defined in Sect.~7 of \citet{PVP_Ale}. The typical uncertainties in \CaFe\ of htese comparison stars are lower than 0.05~dex. This figure shows that 11 of the candidate stars are metal poor (\meta$<$-0.9~dex) and have low \CaFe\ (\CaFe$<$0.3~dex) with respect to the standard halo values, as expected for stars that formed in satellite systems. The upper \meta\ and \CaFe\ boundaries of our selection are defined by the Helmi Stream star $Gaia$ DR3 816615227344979328, which has a confirmed accreted nature in the literature. 
Tab.~\ref{Tab:Accreted} presents the atmospheric parameters and \CaFe\ and \CeFe\ abundances of our finally selected 11 accreted stars. 

Finally, based on this sample of accreted stars, we computed the mean and standard deviation of \meta, \CaFe,\ and \CeFe\ for each system (see  Tab.~\ref{Tab:Accreted}).
        For this sample (although based on low-statistics numbers) and within the error bars, the three accreted systems have a rather similar mean metallicity and extremely close mean [Ca/Fe] values. Moreover,
        Thamnos and GSE appear to have rather similar \CeFe\ values, and
        thus  close values of [Ce/Ca] ratios around $\sim$0.3~dex, which might suggest a rather similar chemical evolution history.
         In contrast, the Helmi Stream appears to be less enriched in cerium, and its [Ce/Ca] ratio is found to be much lower ($\sim$0.07~dex) than for the other two systems. 
        There are very few previous studies on $s$-process abundances in these
        accreted systems. On one hand, \citet{Aguado21} reported a mean barium abundance for GSE lower than our \CeFe\ abundance by about 0.7~dex. On the other hand, \citet{Matsuno21} found some GSE stars enhanced in Ba and La, in agreement with our cerium abundances (with abundances varying from -0.2 to 1.1~dex and mean [Ba/Fe] and [La/Fe] close to 0.4 and 0.2~dex).
        
        Finally, \cite{RecioBlanco2021}, who analysed Y and Eu abundances on Milky Way satellite galaxies, halo stars, and globular clusters, showed that the abundances of another $s$-element [Y/Fe] in low-mass satellites could be slightly lower than in higher-mass satellites in the intermediate-metallicity regime. When a similar behaviour is assumed for Ce,  our lower [Ce/Fe] abundances for the Helmi Stream stars could suggest a lower mass of the parent system of this substructure with respect to the other two. It is interesting to point out that, indeed, \cite{Koppelman19} provided a mass estimate for the Helmi Stream progenitor of about 10$^{8}$ M$_\odot$, while the GES mass estimate from simulations would be six times higher \citep{Helmi20}.
        
        \subsection{Globular clusters: M~4 is Ce enriched}

We also searched for cerium abundances in stars belonging to Galactic globular clusters (GC).
We followed the same procedure as in \citet{PVP_Ale}. First, we cross-matched the Harris catalogue \citep{Harris96} with \gspspec~data ({\it complete sample}) using a maximum separation in the sky of 0.5 degrees.
Then, we rejected all potential GC members whose radial velocity departed
by more than 20~km/s from the median value of each GC.

By this method, we found two stars belonging to M~4 (NGC~6121) ($Gaia$ DR3 6045464990827780608 and 6045463719528135808), and we added two others (6045464166204745344 and 6045490623204749824) found
in \citet{Yong08}, who also studied the two first stars. 
For these four stars, we confirmed that their proper motions agree well with that of M~4, and we found a mean metallicity of -1.20 dex (and a standard 0.08 dex). This mean metallicity is fully compatible with the works of \citet{C09} and \citet{Yong08}, with differences in metallicity  smaller than 0.10 dex, showing the excellent parametrisation of \gspspec~for these stars. \gspspec\ \T~and \g~values of these four stars are also fully compatible with that of \citet{Yong08}. The mean and standard deviation of the differences between the \gspspec\ and \citet{Yong08} values are 1$\pm$45~K and -0.08$\pm$~0.08, respectively.

\gspspec~\CeFe\ abundances also fully agree with those of \citet{Yong08} with a mean difference of -0.05~dex, in the sense \gspspec~minus~\citet{Yong08}, and  a standard deviation of 0.11~dex. 

As a consequence, we found a mean \CeFe\ abundance ratio for M~4 equal to 0.46~$\pm$0.07~dex. This value fully agrees with that found by \citet{Yong08} for their 11 members ([Ce/Fe] = 0.50 $\pm$ 0.10 dex).  We note that \citet{Yong08} also found an enhancement in $s$-process elements (Ce, Ba, and Pb) in M~4 with respect to M~5, a globular cluster whose $s$- and $r$-elements content is similar to that of halo field stars. This could  reveal that the contributing sources of the $s$-process differ between these two globular clusters and may suggest that M~4 could have had a different chemical origin and evolution than M~5 and other halo stars. {We note that our M~4 mean metallicity and [Ce/Fe] abundance are similar to those of Gaia Sausage-Enceladus. The mean [Ca/Fe] value is also similar (0.28 dex).}

Finally, our [Ce/Fe] is also fully compatible (within the error bars) with other $s$-process element abundances reported by \citet{Brown92} from the analysis of three stars ([Ba/Fe] = 0.57 dex and [La/Fe] = 0.43 dex). In summary, our work therefore confirms the enrichment of M4 in $s$-process elements with respect to iron.

\section{Conclusions}
The aim of this paper was to explore the cerium content in the Milky Way disc based on the \Gaia\ \gspspec\ derived chemical abundances. We first validated the LTE \gspspec~cerium abundances with {GALAH DR3}, APOGEE DR17, and \citet{Rebecca19} data. We found a good global agreement, even though {GALAH and }APOGEE do not seem to be on the same scale as \gspspec\ and F19. 

We then selected good-quality samples of \gspspec~cerium abundances using different flag combinations. The general Galactic properties of the selected stars were discussed.  
We found that a majority of these stars are located within $\sim$1~kpc from the Sun, and that the sample is only composed of giant stars. They belong mainly to the disc as more than 95\% of them have a rather low total velocity in the Toomre diagram and a $Z_{\rm max}$ lower than 800~pc. Nevertheless, our sample also contained some metal-poor and cerium-rich stars belonging to the halo, as can be concluded from their velocity, eccentricity, calcium abundances, and their spatial distribution. We also found a young cerium-poor population of stars, as already mentioned in \citet{PVP_Ale}.\\

We studied the chemical evolution of cerium in the Galactic disc. For this purpose, we selected a high-quality sample of stars within the parameter range defined by the most sensitive detection degree of the Ce line we used, and excluding cool AGB stars that might be polluted by internal s-element production. Based on these  different samples, our main results are listed below.

\begin{itemize}

\item The radial and vertical gradients of \CeFe\ and [Ce/H] in the disc were estimated. We found a flat radial gradient in \CeFe\ by adopting both the galactocentric radius and the guiding radius, in agreement with previous findings. The radial gradient in [Ce/H]  is found to be strongly negative, consistent with the radial gradient in metallicity. 
 We also found a strong positive vertical gradient in \CeFe\ and a negative vertical gradient in [Ce/H]. Moreover, we found a slightly increasing [Ce/Ca] versus [Ca/H] up to [Ca/H] < -0.1 dex, showing the later contribution of AGB stars in the Galactic chemical evolution with respect to supernovae II, which are the main producers of $s$-process and $\alpha$-elements, respectively. 

\item Our data can be well reproduced by a new {\it three-infall} Galactic chemical evolution model \citep[see][]{spitoni2022}, in which a timescale for the last gas accretion of about 0.8~Gyr is favoured.

\item Eighty-two stars with Ce abundances belonging to 53 different OCs have been identified. The derived OC \gspspec\ mean metallicities estimated from stars with Ce abundances agree excellenty with the literature metallicities. The relations between OC mean [Ce/Fe], metallicities, and ages were discussed. The derived OC radial gradient in \CeFe\ is compatible with the one derived from field stars (within the error bars). A large proportion of our OCs are very young ($<$ 1 Gyr) and show a large dispersion in cerium abundances.

\end{itemize}

We then explored the Ce content in the Galactic halo. Our results are again listed below.

\begin{itemize}
\item The mean \CeFe\ abundance ratio in the M~4 globular cluster was estimated based on the identification of four of its members. This cluster is found to be enriched in Ce with respect to iron.
\item The cerium abundances in three accreted substructures of the Galactic halo (Helmi Stream, Thamnos, and GSE) were then discussed. Two of the \gspspec\ Helmi Stream stars with derived \CeFe\ were already known in the literature. Their atmospheric parameters as well as their cerium abundance are fully compatible with that derived by \gspspec.  Our sample allowed us to estimate the mean \CeFe\ content in these accreted systems. We found that the Helmi Stream could be slightly underabundant in cerium compared to the two other systems. 

\end{itemize}

All these results confirm the excellent quality of the $Gaia$ data and of the \gspspec~physico-chemical parametrisation.
This study will be extended to the two other $s$-element abundances estimated by \gspspec\ (Nd and Zr), in combination 
with the analysis of Ce-rich AGB stars identified in this work.

\begin{acknowledgements}
    We thank the referee for their valuable comments. ES received funding from the European Union’s Horizon 2020 research and innovation program under SPACE-H2020 grant agreement number 101004214 (EXPLORE project). ARB also acknowledges support from this Horizon program.  PAP and EP 
      thanks the Centre National d'Etudes Spatiales (CNES) for funding support. VG acknowledges support from the European Research Council Consolidator Grant funding scheme (project ASTEROCHRONOMETRY, G.A. n. 772293, http://www.asterochronometry.eu). Special thanks to Niels Nieuwmunster and Botebar for grateful comments on figures.\\
      
      This work has made use of data from the European Space Agency (ESA)
mission \Gaia\ (https://www.cosmos.esa.int/gaia), processed by the \Gaia\ Data Processing and Analysis Consortium (DPAC, https://www.cosmos.
esa.int/web/gaia/dpac/consortium). Funding for the DPAC has been provided by national institutions, in particular the institutions participating in the Gaia Multilateral Agreement.
\end{acknowledgements}

\bibliographystyle{aa} 
\bibliography{ref}

\begin{thebibliography}{96}
\expandafter\ifx\csname natexlab\endcsname\relax\def\natexlab#1{#1}\fi

\bibitem[{{Abdurro'uf} {et~al.}(2022){Abdurro'uf}, {Accetta}, {Aerts}, {Silva
  Aguirre}, {Ahumada}, {Ajgaonkar}, {Filiz Ak}, {Alam}, {Allende Prieto},
  {Almeida}, {Anders}, {Anderson}, {Andrews}, {Anguiano}, {Aquino-Ort{\'\i}z},
  {Arag{\'o}n-Salamanca}, {Argudo-Fern{\'a}ndez}, {Ata}, {Aubert},
  {Avila-Reese}, {Badenes}, {Barb{\'a}}, {Barger}, {Barrera-Ballesteros},
  {Beaton}, {Beers}, {Belfiore}, {Bender}, {Bernardi}, {Bershady}, {Beutler},
  {Bidin}, {Bird}, {Bizyaev}, {Blanc}, {Blanton}, {Boardman}, {Bolton},
  {Boquien}, {Borissova}, {Bovy}, {Brandt}, {Brown}, {Brownstein}, {Brusa},
  {Buchner}, {Bundy}, {Burchett}, {Bureau}, {Burgasser}, {Cabang}, {Campbell},
  {Cappellari}, {Carlberg}, {Wanderley}, {Carrera}, {Cash}, {Chen}, {Chen},
  {Cherinka}, {Chiappini}, {Choi}, {Chojnowski}, {Chung}, {Clerc}, {Cohen},
  {Comerford}, {Comparat}, {da Costa}, {Covey}, {Crane}, {Cruz-Gonzalez},
  {Culhane}, {Cunha}, {Dai}, {Damke}, {Darling}, {Davidson}, {Davies},
  {Dawson}, {De Lee}, {Diamond-Stanic}, {Cano-D{\'\i}az}, {S{\'a}nchez},
  {Donor}, {Duckworth}, {Dwelly}, {Eisenstein}, {Elsworth}, {Emsellem},
  {Eracleous}, {Escoffier}, {Fan}, {Farr}, {Feng}, {Fern{\'a}ndez-Trincado},
  {Feuillet}, {Filipp}, {Fillingham}, {Frinchaboy}, {Fromenteau}, {Galbany},
  {Garc{\'\i}a}, {Garc{\'\i}a-Hern{\'a}ndez}, {Ge}, {Geisler}, {Gelfand},
  {G{\'e}ron}, {Gibson}, {Goddy}, {Godoy-Rivera}, {Grabowski}, {Green},
  {Greener}, {Grier}, {Griffith}, {Guo}, {Guy}, {Hadjara}, {Harding},
  {Hasselquist}, {Hayes}, {Hearty}, {Hern{\'a}ndez}, {Hill}, {Hogg},
  {Holtzman}, {Horta}, {Hsieh}, {Hsu}, {Hsu}, {Huber}, {Huertas-Company},
  {Hutchinson}, {Hwang}, {Ibarra-Medel}, {Chitham}, {Ilha}, {Imig}, {Jaekle},
  {Jayasinghe}, {Ji}, {Johnson}, {Jones}, {J{\"o}nsson}, {Katkov}, {Khalatyan},
  {Kinemuchi}, {Kisku}, {Knapen}, {Kneib}, {Kollmeier}, {Kong}, {Kounkel},
  {Kreckel}, {Krishnarao}, {Lacerna}, {Lane}, {Langgin}, {Lavender}, {Law},
  {Lazarz}, {Leung}, {Leung}, {Lewis}, {Li}, {Li}, {Lian}, {Liang}, {Lin},
  {Lin}, {Lin}, {Lintott}, {Long}, {Longa-Pe{\~n}a}, {L{\'o}pez-Cob{\'a}},
  {Lu}, {Lundgren}, {Luo}, {Mackereth}, {de la Macorra}, {Mahadevan},
  {Majewski}, {Manchado}, {Mandeville}, {Maraston}, {Margalef-Bentabol},
  {Masseron}, {Masters}, {Mathur}, {McDermid}, {Mckay}, {Merloni},
  {Merrifield}, {Meszaros}, {Miglio}, {Di Mille}, {Minniti}, {Minsley},
  {Monachesi}, {Moon}, {Mosser}, {Mulchaey}, {Muna}, {Mu{\~n}oz}, {Myers},
  {Myers}, {Nadathur}, {Nair}, {Nandra}, {Neumann}, {Newman}, {Nidever},
  {Nikakhtar}, {Nitschelm}, {O'Connell}, {Garma-Oehmichen}, {Luan Souza de
  Oliveira}, {Olney}, {Oravetz}, {Ortigoza-Urdaneta}, {Osorio}, {Otter},
  {Pace}, {Padilla}, {Pan}, {Pan}, {Parikh}, {Parker}, {Peirani}, {Pe{\~n}a
  Ram{\'\i}rez}, {Penny}, {Percival}, {Perez-Fournon}, {Pinsonneault},
  {Poidevin}, {Poovelil}, {Price-Whelan}, {B{\'a}rbara de Andrade Queiroz},
  {Raddick}, {Ray}, {Rembold}, {Riddle}, {Riffel}, {Riffel}, {Rix}, {Robin},
  {Rodr{\'\i}guez-Puebla}, {Roman-Lopes}, {Rom{\'a}n-Z{\'u}{\~n}iga}, {Rose},
  {Ross}, {Rossi}, {Rubin}, {Salvato}, {S{\'a}nchez}, {S{\'a}nchez-Gallego},
  {Sanderson}, {Santana Rojas}, {Sarceno}, {Sarmiento}, {Sayres}, {Sazonova},
  {Schaefer}, {Schiavon}, {Schlegel}, {Schneider}, {Schultheis}, {Schwope},
  {Serenelli}, {Serna}, {Shao}, {Shapiro}, {Sharma}, {Shen}, {Shetrone}, {Shu},
  {Simon}, {Skrutskie}, {Smethurst}, {Smith}, {Sobeck}, {Spoo}, {Sprague},
  {Stark}, {Stassun}, {Steinmetz}, {Stello}, {Stone-Martinez},
  {Storchi-Bergmann}, {Stringfellow}, {Stutz}, {Su}, {Taghizadeh-Popp},
  {Talbot}, {Tayar}, {Telles}, {Teske}, {Thakar}, {Theissen}, {Tkachenko},
  {Thomas}, {Tojeiro}, {Hernandez Toledo}, {Troup}, {Trump}, {Trussler},
  {Turner}, {Tuttle}, {Unda-Sanzana}, {V{\'a}zquez-Mata}, {Valentini},
  {Valenzuela}, {Vargas-Gonz{\'a}lez}, {Vargas-Maga{\~n}a}, {Alfaro},
  {Villanova}, {Vincenzo}, {Wake}, {Warfield}, {Washington}, {Weaver},
  {Weijmans}, {Weinberg}, {Weiss}, {Westfall}, {Wild}, {Wilde}, {Wilson},
  {Wilson}, {Wilson}, {Wolf}, {Wood-Vasey}, {Yan}, {Zamora}, {Zasowski},
  {Zhang}, {Zhao}, {Zheng}, {Zheng}, \& {Zhu}}]{APOGEE17}
{Abdurro'uf}, {Accetta}, K., {Aerts}, C., {et~al.} 2022, \apjs, 259, 35

\bibitem[{{Aguado} {et~al.}(2021){Aguado}, {Belokurov}, {Myeong}, {Evans},
  {Kobayashi}, {Sbordone}, {Chanam{\'e}}, {Navarrete}, \& {Koposov}}]{Aguado21}
{Aguado}, D.~S., {Belokurov}, V., {Myeong}, G.~C., {et~al.} 2021, \apjl, 908,
  L8

\bibitem[{{Arlandini} {et~al.}(1999){Arlandini}, {K{\"a}ppeler}, {Wisshak},
  {Gallino}, {Lugaro}, {Busso}, \& {Straniero}}]{Arlandini99}
{Arlandini}, C., {K{\"a}ppeler}, F., {Wisshak}, K., {et~al.} 1999, \apj, 525,
  886

\bibitem[{{Bailer-Jones} {et~al.}(2021){Bailer-Jones}, {Rybizki}, {Fouesneau},
  {Demleitner}, \& {Andrae}}]{Coryn21}
{Bailer-Jones}, C.~A.~L., {Rybizki}, J., {Fouesneau}, M., {Demleitner}, M., \&
  {Andrae}, R. 2021, \aj, 161, 147

\bibitem[{{Battistini} \& {Bensby}(2016)}]{BB16}
{Battistini}, C. \& {Bensby}, T. 2016, \aap, 586, A49

\bibitem[{{Belokurov} {et~al.}(2018){Belokurov}, {Erkal}, {Evans}, {Koposov},
  \& {Deason}}]{Belokurov18}
{Belokurov}, V., {Erkal}, D., {Evans}, N.~W., {Koposov}, S.~E., \& {Deason},
  A.~J. 2018, \mnras, 478, 611

\bibitem[{{Bernard}(2017)}]{bernard2017}
{Bernard}, E.~J. 2017, in SF2A-2017: Proceedings of the Annual meeting of the
  French Society of Astronomy and Astrophysics, ed. C.~{Reyl{\'e}}, P.~{Di
  Matteo}, F.~{Herpin}, E.~{Lagadec}, A.~{Lan{\c{c}}on}, Z.~{Meliani}, \&
  F.~{Royer}, Di

\bibitem[{{Bijaoui}(2012)}]{2012ada..confE...2B}
{Bijaoui}, A. 2012, in Seventh Conference on Astronomical Data Analysis, ed.
  J.-L. {Starck} \& C.~{Surace}, 2

\bibitem[{{Birch} \& {Downs}(1994)}]{1994Metro..31..315B}
{Birch}, K.~P. \& {Downs}, M.~J. 1994, Metrologia, 31, 315

\bibitem[{{Bisterzo} {et~al.}(2015){Bisterzo}, {Gallino}, {K{\"a}ppeler},
  {Wiescher}, {Imbriani}, {Straniero}, {Cristallo}, {G{\"o}rres}, \&
  {deBoer}}]{Bisterzo15}
{Bisterzo}, S., {Gallino}, R., {K{\"a}ppeler}, F., {et~al.} 2015, \mnras, 449,
  506

\bibitem[{{Bisterzo} {et~al.}(2011){Bisterzo}, {Gallino}, {Straniero},
  {Cristallo}, \& {K{\"a}ppeler}}]{Bisterzo11}
{Bisterzo}, S., {Gallino}, R., {Straniero}, O., {Cristallo}, S., \&
  {K{\"a}ppeler}, F. 2011, \mnras, 418, 284

\bibitem[{{Bisterzo} {et~al.}(2016){Bisterzo}, {Travaglio}, {Wiescher},
  {Gallino}, {K{\"o}ppeler}, {Straniero}, {Cristallo}, {Imbriani},
  {G{\"o}rres}, \& {deBoer}}]{Bisterzo16}
{Bisterzo}, S., {Travaglio}, C., {Wiescher}, M., {et~al.} 2016, in Journal of
  Physics Conference Series, Vol. 665, Journal of Physics Conference Series,
  012023

\bibitem[{{Brown} \& {Wallerstein}(1992)}]{Brown92}
{Brown}, J.~A. \& {Wallerstein}, G. 1992, \aj, 104, 1818

\bibitem[{{Buder} {et~al.}(2021){Buder}, {Sharma}, {Kos}, {Amarsi},
  {Nordlander}, {Lind}, {Martell}, {Asplund}, {Bland-Hawthorn}, {Casey}, {de
  Silva}, {D'Orazi}, {Freeman}, {Hayden}, {Lewis}, {Lin}, {Schlesinger},
  {Simpson}, {Stello}, {Zucker}, {Zwitter}, {Beeson}, {Buck}, {Casagrande},
  {Clark}, {{\v{C}}otar}, {da Costa}, {de Grijs}, {Feuillet}, {Horner},
  {Kafle}, {Khanna}, {Kobayashi}, {Liu}, {Montet}, {Nandakumar}, {Nataf},
  {Ness}, {Spina}, {Tepper-Garc{\'\i}a}, {Ting}, {Traven},
  {Vogrin{\v{c}}i{\v{c}}}, {Wittenmyer}, {Wyse}, {{\v{Z}}erjal}, \& {Galah
  Collaboration}}]{GALAHDR3}
{Buder}, S., {Sharma}, S., {Kos}, J., {et~al.} 2021, \mnras, 506, 150

\bibitem[{{Burbidge} {et~al.}(1957){Burbidge}, {Burbidge}, {Fowler}, \&
  {Hoyle}}]{Burbidge57}
{Burbidge}, E.~M., {Burbidge}, G.~R., {Fowler}, W.~A., \& {Hoyle}, F. 1957,
  Reviews of Modern Physics, 29, 547

\bibitem[{{Burris} {et~al.}(2000){Burris}, {Pilachowski}, {Armandroff},
  {Sneden}, {Cowan}, \& {Roe}}]{Burris00}
{Burris}, D.~L., {Pilachowski}, C.~A., {Armandroff}, T.~E., {et~al.} 2000,
  \apj, 544, 302

\bibitem[{{Busso} {et~al.}(1999){Busso}, {Gallino}, \& {Wasserburg}}]{Busso99}
{Busso}, M., {Gallino}, R., \& {Wasserburg}, G.~J. 1999, \araa, 37, 239

\bibitem[{{Cantat-Gaudin} {et~al.}(2020){Cantat-Gaudin}, {Anders},
  {Castro-Ginard}, {Jordi}, {Romero-G{\'o}mez}, {Soubiran}, {Casamiquela},
  {Tarricq}, {Moitinho}, {Vallenari}, {Bragaglia}, {Krone-Martins}, \&
  {Kounkel}}]{CG20}
{Cantat-Gaudin}, T., {Anders}, F., {Castro-Ginard}, A., {et~al.} 2020, \aap,
  640, A1

\bibitem[{{Carretta} {et~al.}(2009){Carretta}, {Bragaglia}, {Gratton},
  {D'Orazi}, \& {Lucatello}}]{C09}
{Carretta}, E., {Bragaglia}, A., {Gratton}, R., {D'Orazi}, V., \& {Lucatello},
  S. 2009, \aap, 508, 695

\bibitem[{{Casamiquela} {et~al.}(2021){Casamiquela}, {Soubiran}, {Jofr{\'e}},
  {Chiappini}, {Lagarde}, {Tarricq}, {Carrera}, {Jordi},
  {Balaguer-N{\'u}{\~n}ez}, {Carbajo-Hijarrubia}, \&
  {Blanco-Cuaresma}}]{Casamiquela21}
{Casamiquela}, L., {Soubiran}, C., {Jofr{\'e}}, P., {et~al.} 2021, \aap, 652,
  A25

\bibitem[{{Castro-Ginard} {et~al.}(2022){Castro-Ginard}, {Jordi}, {Luri},
  {Cantat-Gaudin}, {Carrasco}, {Casamiquela}, {Anders},
  {Balaguer-N{\'u}{\~n}ez}, \& {Badia}}]{CG21}
{Castro-Ginard}, A., {Jordi}, C., {Luri}, X., {et~al.} 2022, \aap, 661, A118

\bibitem[{{Cescutti} \& {Chiappini}(2014)}]{cescutti2014}
{Cescutti}, G. \& {Chiappini}, C. 2014, \aap, 565, A51

\bibitem[{{Cescutti} {et~al.}(2013){Cescutti}, {Chiappini}, {Hirschi},
  {Meynet}, \& {Frischknecht}}]{cescutti2013}
{Cescutti}, G., {Chiappini}, C., {Hirschi}, R., {Meynet}, G., \&
  {Frischknecht}, U. 2013, \aap, 553, A51

\bibitem[{{Cescutti} {et~al.}(2015){Cescutti}, {Romano}, {Matteucci},
  {Chiappini}, \& {Hirschi}}]{cescutti2015}
{Cescutti}, G., {Romano}, D., {Matteucci}, F., {Chiappini}, C., \& {Hirschi},
  R. 2015, \aap, 577, A139

\bibitem[{{Clayton} \& {Rassbach}(1967)}]{CR67}
{Clayton}, D.~D. \& {Rassbach}, M.~E. 1967, \apj, 148, 69

\bibitem[{{Contursi} {et~al.}(2021){Contursi}, {de Laverny}, {Recio-Blanco}, \&
  {Palicio}}]{BestArticleEver}
{Contursi}, G., {de Laverny}, P., {Recio-Blanco}, A., \& {Palicio}, P.~A. 2021,
  \aap, 654, A130

\bibitem[{{Cristallo} {et~al.}(2011){Cristallo}, {Piersanti}, {Straniero},
  {Gallino}, {Dom{\'\i}nguez}, {Abia}, {Di Rico}, {Quintini}, \&
  {Bisterzo}}]{cristallo2011}
{Cristallo}, S., {Piersanti}, L., {Straniero}, O., {et~al.} 2011, \apjs, 197,
  17

\bibitem[{{Cristallo} {et~al.}(2009){Cristallo}, {Straniero}, {Gallino},
  {Piersanti}, {Dom{\'\i}nguez}, \& {Lederer}}]{cristallo2009}
{Cristallo}, S., {Straniero}, O., {Gallino}, R., {et~al.} 2009, \apj, 696, 797

\bibitem[{{Cropper} {et~al.}(2018){Cropper}, {Katz}, {Sartoretti}, {Prusti},
  {de Bruijne}, {Chassat}, {Charvet}, {Boyadjian}, {Perryman}, {Sarri}, {Gare},
  {Erdmann}, {Munari}, {Zwitter}, {Wilkinson}, {Arenou}, {Vallenari},
  {G{\'o}mez}, {Panuzzo}, {Seabroke}, {Allende Prieto}, {Benson}, {Marchal},
  {Huckle}, {Smith}, {Dolding}, {Jan{\ss}en}, {Viala}, {Blomme}, {Baker},
  {Boudreault}, {Crifo}, {Soubiran}, {Fr{\'e}mat}, {Jasniewicz}, {Guerrier},
  {Guy}, {Turon}, {Jean-Antoine-Piccolo}, {Th{\'e}venin}, {David}, {Gosset}, \&
  {Damerdji}}]{RVS18}
{Cropper}, M., {Katz}, D., {Sartoretti}, P., {et~al.} 2018, \aap, 616, A5

\bibitem[{{da Silva} {et~al.}(2016){da Silva}, {Lemasle}, {Bono}, {Genovali},
  {McWilliam}, {Cristallo}, {Bergemann}, {Buonanno}, {Fabrizio}, {Ferraro},
  {Fran{\c{c}}ois}, {Iannicola}, {Inno}, {Laney}, {Kudritzki}, {Matsunaga},
  {Nonino}, {Primas}, {Przybilla}, {Romaniello}, {Th{\'e}venin}, \&
  {Urbaneja}}]{DaSilva16}
{da Silva}, R., {Lemasle}, B., {Bono}, G., {et~al.} 2016, \aap, 586, A125

\bibitem[{{De Cia} {et~al.}(2021){De Cia}, {Jenkins}, {Fox}, {Ledoux},
  {Ramburuth-Hurt}, {Konstantopoulou}, {Petitjean}, \& {Krogager}}]{DeCia21}
{De Cia}, A., {Jenkins}, E.~B., {Fox}, A.~J., {et~al.} 2021, \nat, 597, 206

\bibitem[{{Delgado Mena} {et~al.}(2017){Delgado Mena}, {Tsantaki}, {Adibekyan},
  {Sousa}, {Santos}, {Gonz{\'a}lez Hern{\'a}ndez}, \& {Israelian}}]{DM17}
{Delgado Mena}, E., {Tsantaki}, M., {Adibekyan}, V.~Z., {et~al.} 2017, \aap,
  606, A94

\bibitem[{{Feuillet} {et~al.}(2020){Feuillet}, {Feltzing}, {Sahlholdt}, \&
  {Casagrande}}]{Feuillet20}
{Feuillet}, D.~K., {Feltzing}, S., {Sahlholdt}, C.~L., \& {Casagrande}, L.
  2020, \mnras, 497, 109

\bibitem[{{Feuillet} {et~al.}(2021){Feuillet}, {Sahlholdt}, {Feltzing}, \&
  {Casagrande}}]{Feuillet21}
{Feuillet}, D.~K., {Sahlholdt}, C.~L., {Feltzing}, S., \& {Casagrande}, L.
  2021, \mnras, 508, 1489

\bibitem[{{Forsberg} {et~al.}(2019){Forsberg}, {J{\"o}nsson}, {Ryde}, \&
  {Matteucci}}]{Rebecca19}
{Forsberg}, R., {J{\"o}nsson}, H., {Ryde}, N., \& {Matteucci}, F. 2019, \aap,
  631, A113

\bibitem[{{Freiburghaus} {et~al.}(1999){Freiburghaus}, {Rosswog}, \&
  {Thielemann}}]{Freiburghaus99}
{Freiburghaus}, C., {Rosswog}, S., \& {Thielemann}, F.~K. 1999, \apjl, 525,
  L121

\bibitem[{{Frischknecht} {et~al.}(2016){Frischknecht}, {Hirschi}, {Pignatari},
  {Maeder}, {Meynet}, {Chiappini}, {Thielemann}, {Rauscher}, {Georgy}, \&
  {Ekstr{\"o}m}}]{frisch2016}
{Frischknecht}, U., {Hirschi}, R., {Pignatari}, M., {et~al.} 2016, \mnras, 456,
  1803

\bibitem[{{Frischknecht} {et~al.}(2012){Frischknecht}, {Hirschi}, \&
  {Thielemann}}]{frisck2012}
{Frischknecht}, U., {Hirschi}, R., \& {Thielemann}, F.~K. 2012, \aap, 538, L2

\bibitem[{{Gaia Collaboration, Recio-Blanco} {et~al.}(2022){Gaia Collaboration,
  Recio-Blanco}, {Kordopatis}, {de Laverny}, {Palicio}, {Spagna}, {Spina},
  {Katz}, {Re Fiorentin}, {Poggio}, {McMillan}, {Vallenari}, {Lattanzi},
  {Seabroke}, {Casamiquela}, {Bragaglia}, {Antoja}, {Bailer-Jones}, {Andrae},
  {Fouesneau}, {Cropper}, {Cantat-Gaudin}, {Heiter}, {Bijaoui}, {Brown},
  {Prusti}, {de Bruijne}, {Arenou}, {Babusiaux}, {Biermann}, {Creevey},
  {Ducourant}, {Evans}, {Eyer}, {Guerra}, {Hutton}, {Jordi}, {Klioner},
  {Lammers}, {Lindegren}, {Luri}, {Mignard}, {Panem}, {Pourbaix}, {Randich},
  {Sartoretti}, {Soubiran}, {Tanga}, {Walton}, {Bastian}, {Drimmel}, {Jansen},
  {van Leeuwen}, {Bakker}, {Cacciari}, {Casta{\~n}eda}, {De Angeli},
  {Fabricius}, {Fr{\'e}mat}, {Galluccio}, {Guerrier}, {Masana}, {Messineo},
  {Mowlavi}, {Nicolas}, {Nienartowicz}, {Pailler}, {Panuzzo}, {Riclet}, {Roux},
  {Sordo}, {Th{\'e}venin}, {Gracia-Abril}, {Portell}, {Teyssier}, {Altmann},
  {Audard}, {Bellas-Velidis}, {Benson}, {Berthier}, {Blomme}, {Burgess},
  {Busonero}, {Busso}, {C{\'a}novas}, {Carry}, {Cellino}, {Cheek},
  {Clementini}, {Damerdji}, {Davidson}, {de Teodoro}, {Nu{\~n}ez Campos},
  {Delchambre}, {Dell'Oro}, {Esquej}, {Fern{\'a}ndez-Hern{\'a}ndez}, {Fraile},
  {Garabato}, {Garc{\'\i}a-Lario}, {Gosset}, {Haigron}, {Halbwachs}, {Hambly},
  {Harrison}, {Hern{\'a}ndez}, {Hestroffer}, {Hodgkin}, {Holl}, {Jan{\ss}en},
  {Jevardat de Fombelle}, {Jordan}, {Krone-Martins}, {Lanzafame},
  {L{\"o}ffler}, {Marchal}, {Marrese}, {Moitinho}, {Muinonen}, {Osborne},
  {Pancino}, {Pauwels}, {Reyl{\'e}}, {Riello}, {Rimoldini}, {Roegiers},
  {Rybizki}, {Sarro}, {Siopis}, {Smith}, {Sozzetti}, {Utrilla}, {van Leeuwen},
  {Abbas}, {{\'A}brah{\'a}m}, {Abreu Aramburu}, {Aerts}, {Aguado}, {Ajaj},
  {Aldea-Montero}, {Altavilla}, {{\'A}lvarez}, {Alves}, {Anders}, {Anderson},
  {Anglada Varela}, {Baines}, {Baker}, {Balaguer-N{\'u}{\~n}ez}, {Balbinot},
  {Balog}, {Barache}, {Barbato}, {Barros}, {Barstow}, {Bartolom{\'e}},
  {Bassilana}, {Bauchet}, {Becciani}, {Bellazzini}, {Berihuete}, {Bernet},
  {Bertone}, {Bianchi}, {Binnenfeld}, {Blanco-Cuaresma}, {Boch}, {Bombrun},
  {Bossini}, {Bouquillon}, {Bramante}, {Breedt}, {Bressan}, {Brouillet},
  {Brugaletta}, {Bucciarelli}, {Burlacu}, {Butkevich}, {Buzzi}, {Caffau},
  {Cancelliere}, {Carballo}, {Carlucci}, {Carnerero}, {Carrasco}, {Castellani},
  {Castro-Ginard}, {Chaoul}, {Charlot}, {Chemin}, {Chiaramida}, {Chiavassa},
  {Chornay}, {Comoretto}, {Contursi}, {Cooper}, {Cornez}, {Cowell}, {Crifo},
  {Crosta}, {Crowley}, {Dafonte}, {Dapergolas}, {David}, {De Luise}, {De
  March}, {De Ridder}, {de Souza}, {de Torres}, {del Peloso}, {del Pozo},
  {Delbo}, {Delgado}, {Delisle}, {Demouchy}, {Dharmawardena}, {Di Matteo},
  {Diakite}, {Diener}, {Distefano}, {Dolding}, {Edvardsson}, {Enke}, {Fabre},
  {Fabrizio}, {Faigler}, {Fedorets}, {Fernique}, {Figueras}, {Fournier},
  {Fouron}, {Fragkoudi}, {Gai}, {Garcia-Gutierrez}, {Garcia-Reinaldos},
  {Garc{\'\i}a-Torres}, {Garofalo}, {Gavel}, {Gavras}, {Gerlach}, {Geyer},
  {Giacobbe}, {Gilmore}, {Girona}, {Giuffrida}, {Gomel}, {Gomez},
  {Gonz{\'a}lez-N{\'u}{\~n}ez}, {Gonz{\'a}lez-Santamar{\'\i}a},
  {Gonz{\'a}lez-Vidal}, {Granvik}, {Guillout}, {Guiraud},
  {Guti{\'e}rrez-S{\'a}nchez}, {Guy}, {Hatzidimitriou}, {Hauser}, {Haywood},
  {Helmer}, {Helmi}, {Sarmiento}, {Hidalgo}, {H{\l}adczuk}, {Hobbs}, {Holland},
  {Huckle}, {Jardine}, {Jasniewicz}, {Jean-Antoine Piccolo},
  {Jim{\'e}nez-Arranz}, {Juaristi Campillo}, {Julbe}, {Karbevska}, {Kervella},
  {Khanna}, {Korn}, {K{\'o}sp{\'a}l}, {Kostrzewa-Rutkowska}, {Kruszy{\'n}ska},
  {Kun}, {Laizeau}, {Lambert}, {Lanza}, {Lasne}, {Le Campion}, {Lebreton},
  {Lebzelter}, {Leccia}, {Leclerc}, {Lecoeur-Taibi}, {Liao}, {Licata},
  {Lindstr{\o}m}, {Lister}, {Livanou}, {Lobel}, {Lorca}, {Loup}, {Madrero
  Pardo}, {Magdaleno Romeo}, {Managau}, {Mann}, {Manteiga}, {Marchant},
  {Marconi}, {Marcos}, {Marcos Santos}, {Mar{\'\i}n Pina}, {Marinoni},
  {Marocco}, {Marshall}, {Polo}, {Mart{\'\i}n-Fleitas}, {Marton}, {Mary},
  {Masip}, {Massari}, {Mastrobuono-Battisti}, {Mazeh}, {Messina}, {Michalik},
  {Millar}, {Mints}, {Molina}, {Molinaro}, {Moln{\'a}r}, {Monari},
  {Mongui{\'o}}, {Montegriffo}, {Montero}, {Mor}, {Mora}, {Morbidelli},
  {Morel}, {Morris}, {Muraveva}, {Murphy}, {Musella}, {Nagy}, {Noval},
  {Oca{\~n}a}, {Ogden}, {Ordenovic}, {Osinde}, {Pagani}, {Pagano}, {Palaversa},
  {Pallas-Quintela}, {Panahi}, {Payne-Wardenaar}, {Pe{\~n}alosa Esteller},
  {Penttil{\"a}}, {Pichon}, {Piersimoni}, {Pineau}, {Plachy}, {Plum},
  {Pr{\v{s}}a}, {Pulone}, {Racero}, {Ragaini}, {Rainer}, {Raiteri}, {Ramos},
  {Ramos-Lerate}, {Regibo}, {Richards}, {Rios Diaz}, {Ripepi}, {Riva}, {Rix},
  {Rixon}, {Robichon}, {Robin}, {Robin}, {Roelens}, {Rogues}, {Rohrbasser},
  {Romero-G{\'o}mez}, {Rowell}, {Royer}, {Ruz Mieres}, {Rybicki}, {Sadowski},
  {S{\'a}ez N{\'u}{\~n}ez}, {Sagrist{\`a} Sell{\'e}s}, {Sahlmann}, {Salguero},
  {Samaras}, {Sanchez Gimenez}, {Sanna}, {Santove{\~n}a}, {Sarasso},
  {Schultheis}, {Sciacca}, {Segol}, {Segovia}, {S{\'e}gransan}, {Semeux},
  {Shahaf}, {Siddiqui}, {Siebert}, {Siltala}, {Silvelo}, {Slezak}, {Slezak},
  {Smart}, {Snaith}, {Solano}, {Solitro}, {Souami}, {Souchay}, {Spoto},
  {Steele}, {Steidelm{\"u}ller}, {Stephenson}, {S{\"u}veges}, {Surdej},
  {Szabados}, {Szegedi-Elek}, {Taris}, {Taylor}, {Teixeira}, {Tolomei},
  {Tonello}, {Torra}, {Torra}, {Torralba Elipe}, {Trabucchi}, {Tsounis},
  {Turon}, {Ulla}, {Unger}, {Vaillant}, {van Dillen}, {van Reeven}, {Vanel},
  {Vecchiato}, {Viala}, {Vicente}, {Voutsinas}, {Weiler}, {Wevers},
  {Wyrzykowski}, {Yoldas}, {Yvard}, {Zhao}, {Zorec}, {Zucker}, \&
  {Zwitter}}]{PVP_Ale}
{Gaia Collaboration, Recio-Blanco}, A., {Kordopatis}, G., {de Laverny}, P.,
  {et~al.} 2022, arXiv e-prints, arXiv:2206.05534

\bibitem[{{Gaia Collaboration, Vallenari} {et~al.}(2022){Gaia Collaboration,
  Vallenari}, {A.}, {Brown, A.G.A.}, {Prusti, T.}, \& {et al.}}]{Vallenari22}
{Gaia Collaboration, Vallenari}, {A.}, {Brown, A.G.A.}, {Prusti, T.}, \& {et
  al.} 2022, A\&A

\bibitem[{{Gallino} {et~al.}(1998){Gallino}, {Arlandini}, {Busso}, {Lugaro},
  {Travaglio}, {Straniero}, {Chieffi}, \& {Limongi}}]{Gallino98}
{Gallino}, R., {Arlandini}, C., {Busso}, M., {et~al.} 1998, \apj, 497, 388

\bibitem[{{Grevesse} {et~al.}(2007){Grevesse}, {Asplund}, \&
  {Sauval}}]{Grevesse07}
{Grevesse}, N., {Asplund}, M., \& {Sauval}, A.~J. 2007, \ssr, 130, 105

\bibitem[{{Grevesse} {et~al.}(2015){Grevesse}, {Scott}, {Asplund}, \&
  {Sauval}}]{Grevesse15}
{Grevesse}, N., {Scott}, P., {Asplund}, M., \& {Sauval}, A.~J. 2015, \aap, 573,
  A27

\bibitem[{{Griffith} {et~al.}(2021){Griffith}, {Weinberg}, {Johnson}, {Beaton},
  {Garc{\'\i}a-Hern{\'a}ndez}, {Hasselquist}, {Holtzman}, {Johnson},
  {J{\"o}nsson}, {Lane}, {Nataf}, \& {Roman-Lopes}}]{Griffith21}
{Griffith}, E., {Weinberg}, D.~H., {Johnson}, J.~A., {et~al.} 2021, \apj, 909,
  77

\bibitem[{{Grisoni} {et~al.}(2020){Grisoni}, {Cescutti}, {Matteucci},
  {Forsberg}, {J{\"o}nsson}, \& {Ryde}}]{grisoni2020}
{Grisoni}, V., {Cescutti}, G., {Matteucci}, F., {et~al.} 2020, \mnras, 492,
  2828

\bibitem[{{Grisoni} {et~al.}(2017){Grisoni}, {Spitoni}, {Matteucci},
  {Recio-Blanco}, {de Laverny}, {Hayden}, {Mikolaitis}, \&
  {Worley}}]{grisoni2017}
{Grisoni}, V., {Spitoni}, E., {Matteucci}, F., {et~al.} 2017, \mnras, 472, 3637

\bibitem[{{Gull} {et~al.}(2021){Gull}, {Frebel}, {Hinojosa}, {Roederer}, {Ji},
  \& {Brauer}}]{Gull21}
{Gull}, M., {Frebel}, A., {Hinojosa}, K., {et~al.} 2021, \apj, 912, 52

\bibitem[{{Harris}(1996)}]{Harris96}
{Harris}, W.~E. 1996, \aj, 112, 1487

\bibitem[{{Helmi}(2020)}]{Helmi20}
{Helmi}, A. 2020, \araa, 58, 205

\bibitem[{{Helmi} {et~al.}(2018){Helmi}, {Babusiaux}, {Koppelman}, {Massari},
  {Veljanoski}, \& {Brown}}]{Helmi18}
{Helmi}, A., {Babusiaux}, C., {Koppelman}, H.~H., {et~al.} 2018, \nat, 563, 85

\bibitem[{{Helmi} {et~al.}(1999){Helmi}, {White}, {de Zeeuw}, \&
  {Zhao}}]{Helmi99}
{Helmi}, A., {White}, S. D.~M., {de Zeeuw}, P.~T., \& {Zhao}, H. 1999, \nat,
  402, 53

\bibitem[{{Ishigaki} {et~al.}(2013){Ishigaki}, {Aoki}, \& {Chiba}}]{Ishigaki13}
{Ishigaki}, M.~N., {Aoki}, W., \& {Chiba}, M. 2013, \apj, 771, 67

\bibitem[{{Kappeler} {et~al.}(1989){Kappeler}, {Beer}, \&
  {Wisshak}}]{Kappeler89}
{Kappeler}, F., {Beer}, H., \& {Wisshak}, K. 1989, Reports on Progress in
  Physics, 52, 945

\bibitem[{{Karakas} \& {Lattanzio}(2014)}]{KL14}
{Karakas}, A.~I. \& {Lattanzio}, J.~C. 2014, \pasa, 31, e030

\bibitem[{{Katz} {et~al.}(2022){Katz}, {Sartoretti}, {Guerrier}, {Panuzzo},
  {Seabroke}, {Th{\'e}venin}, {Cropper}, {Benson}, {Blomme}, {Haigron},
  {Marchal}, {Smith}, {Baker}, {Chemin}, {Damerdji}, {David}, {Dolding},
  {Fr{\'e}mat}, {Gosset}, {Jan{\ss}en}, {Jasniewicz}, {Lobel}, {Plum},
  {Samaras}, {Snaith}, {Soubiran}, {Vanel}, {Zwitter}, {Antoja}, {Arenou},
  {Babusiaux}, {Brouillet}, {Caffau}, {Di Matteo}, {Fabre}, {Fabricius},
  {Frakgoudi}, {Haywood}, {Huckle}, {Hottier}, {Lasne}, {Leclerc},
  {Mastrobuono-Battisti}, {Royer}, {Teyssier}, {Zorec}, {Crifo}, {Jean-Antoine
  Piccolo}, {Turon}, \& {Viala}}]{Katz22}
{Katz}, D., {Sartoretti}, P., {Guerrier}, A., {et~al.} 2022, arXiv e-prints,
  arXiv:2206.05902

\bibitem[{{Koppelman} {et~al.}(2019){Koppelman}, {Helmi}, {Massari},
  {Price-Whelan}, \& {Starkenburg}}]{Koppelman19}
{Koppelman}, H.~H., {Helmi}, A., {Massari}, D., {Price-Whelan}, A.~M., \&
  {Starkenburg}, T.~K. 2019, \aap, 631, L9

\bibitem[{{Lamb} {et~al.}(1977){Lamb}, {Howard}, {Truran}, \& {Iben}}]{Lamb77}
{Lamb}, S.~A., {Howard}, W.~M., {Truran}, J.~W., \& {Iben}, I., J. 1977, \apj,
  217, 213

\bibitem[{{Lawler} {et~al.}(2009){Lawler}, {Sneden}, {Cowan}, {Ivans}, \& {Den
  Hartog}}]{Lawler09}
{Lawler}, J.~E., {Sneden}, C., {Cowan}, J.~J., {Ivans}, I.~I., \& {Den Hartog},
  E.~A. 2009, \apjs, 182, 51

\bibitem[{{Limongi} \& {Chieffi}(2018)}]{limongi2018}
{Limongi}, M. \& {Chieffi}, A. 2018, \apjs, 237, 13

\bibitem[{{Magrini} {et~al.}(2018){Magrini}, {Spina}, {Randich}, {Friel},
  {Kordopatis}, {Worley}, {Pancino}, {Bragaglia}, {Donati},
  {Tautvai{\v{s}}ien{\.{e}}}, {Bagdonas}, {Delgado-Mena}, {Adibekyan}, {Sousa},
  {Jim{\'e}nez-Esteban}, {Sanna}, {Roccatagliata}, {Bonito}, {Sbordone},
  {Duffau}, {Gilmore}, {Feltzing}, {Jeffries}, {Vallenari}, {Alfaro}, {Bensby},
  {Francois}, {Koposov}, {Korn}, {Recio-Blanco}, {Smiljanic}, {Bayo},
  {Carraro}, {Casey}, {Costado}, {Damiani}, {Franciosini}, {Frasca},
  {Hourihane}, {Jofr{\'e}}, {de Laverny}, {Lewis}, {Masseron}, {Monaco},
  {Morbidelli}, {Prisinzano}, {Sacco}, \& {Zaggia}}]{Magrini18}
{Magrini}, L., {Spina}, L., {Randich}, S., {et~al.} 2018, \aap, 617, A106

\bibitem[{{Maiorca} {et~al.}(2011){Maiorca}, {Randich}, {Busso}, {Magrini}, \&
  {Palmerini}}]{Maiorca11}
{Maiorca}, E., {Randich}, S., {Busso}, M., {Magrini}, L., \& {Palmerini}, S.
  2011, \apj, 736, 120

\bibitem[{{Mashonkina} {et~al.}(2007){Mashonkina}, {Vinogradova}, {Ptitsyn},
  {Khokhlova}, \& {Chernetsova}}]{Mashonkina07}
{Mashonkina}, L.~I., {Vinogradova}, A.~B., {Ptitsyn}, D.~A., {Khokhlova},
  V.~S., \& {Chernetsova}, T.~A. 2007, Astronomy Reports, 51, 903

\bibitem[{{Matsuno} {et~al.}(2021){Matsuno}, {Hirai}, {Tarumi}, {Hotokezaka},
  {Tanaka}, \& {Helmi}}]{Matsuno21}
{Matsuno}, T., {Hirai}, Y., {Tarumi}, Y., {et~al.} 2021, \aap, 650, A110

\bibitem[{{Matteucci} {et~al.}(2014){Matteucci}, {Romano}, {Arcones},
  {Korobkin}, \& {Rosswog}}]{matteucci2014}
{Matteucci}, F., {Romano}, D., {Arcones}, A., {Korobkin}, O., \& {Rosswog}, S.
  2014, \mnras, 438, 2177

\bibitem[{{Mishenina} \& {Kovtyukh}(2001)}]{Mishenina01}
{Mishenina}, T.~V. \& {Kovtyukh}, V.~V. 2001, \aap, 370, 951

\bibitem[{{Mishenina} {et~al.}(2013){Mishenina}, {Pignatari}, {Korotin},
  {Soubiran}, {Charbonnel}, {Thielemann}, {Gorbaneva}, \&
  {Basak}}]{Mishenina13}
{Mishenina}, T.~V., {Pignatari}, M., {Korotin}, S.~A., {et~al.} 2013, \aap,
  552, A128

\bibitem[{{Myeong} {et~al.}(2018){Myeong}, {Evans}, {Belokurov}, {Amorisco}, \&
  {Koposov}}]{Myeong18}
{Myeong}, G.~C., {Evans}, N.~W., {Belokurov}, V., {Amorisco}, N.~C., \&
  {Koposov}, S.~E. 2018, \mnras, 475, 1537

\bibitem[{{Nishimura} {et~al.}(2006){Nishimura}, {Kotake}, {Hashimoto},
  {Yamada}, {Nishimura}, {Fujimoto}, \& {Sato}}]{Nishimura06}
{Nishimura}, S., {Kotake}, K., {Hashimoto}, M.-a., {et~al.} 2006, \apj, 642,
  410

\bibitem[{{Palicio} {et~al.}(2022){Palicio}, {Recio-Blanco}, {Poggio},
  {Antoja}, {McMillan}, \& {Spitoni}}]{Pedro}
{Palicio}, P.~A., {Recio-Blanco}, A., {Poggio}, E., {et~al.} 2022, arXiv
  e-prints, arXiv:2209.09989

\bibitem[{{Peters}(1968)}]{Peters68}
{Peters}, J.~G. 1968, \apj, 154, 225

\bibitem[{{Pignatari} {et~al.}(2010){Pignatari}, {Gallino}, {Heil}, {Wiescher},
  {K{\"a}ppeler}, {Herwig}, \& {Bisterzo}}]{Pignatari10}
{Pignatari}, M., {Gallino}, R., {Heil}, M., {et~al.} 2010, \apj, 710, 1557

\bibitem[{{Poggio} {et~al.}(2022){Poggio}, {Recio-Blanco}, {Palicio}, {Re
  Fiorentin}, {de Laverny}, {Drimmel}, {Kordopatis}, {Lattanzi}, {Schultheis},
  {Spagna}, \& {Spitoni}}]{Poggio22}
{Poggio}, E., {Recio-Blanco}, A., {Palicio}, P.~A., {et~al.} 2022, \aap, 666,
  L4

\bibitem[{{Prantzos} {et~al.}(2018){Prantzos}, {Abia}, {Limongi}, {Chieffi}, \&
  {Cristallo}}]{Nikos18}
{Prantzos}, N., {Abia}, C., {Limongi}, M., {Chieffi}, A., \& {Cristallo}, S.
  2018, \mnras, 476, 3432

\bibitem[{{Recio-Blanco} {et~al.}(2016){Recio-Blanco}, {de Laverny}, {Allende
  Prieto}, {Fustes}, {Manteiga}, {Arcay}, {Bijaoui}, {Dafonte}, {Ordenovic}, \&
  {Ordo{\~n}ez Blanco}}]{RB16}
{Recio-Blanco}, A., {de Laverny}, P., {Allende Prieto}, C., {et~al.} 2016,
  \aap, 585, A93

\bibitem[{{Recio-Blanco} {et~al.}(2022){Recio-Blanco}, {de Laverny}, {Palicio},
  {Kordopatis}, {{\'A}lvarez}, {Schultheis}, {Contursi}, {Zhao}, {Torralba
  Elipe}, {Ordenovic}, {Manteiga}, {Dafonte}, {Oreshina-Slezak}, {Bijaoui},
  {Fremat}, {Seabroke}, {Pailler}, {Spitoni}, {Poggio}, {Creevey}, {Abreu
  Aramburu}, {Accart}, {Andrae}, {Bailer-Jones}, {Bellas-Velidis}, {Brouillet},
  {Brugaletta}, {Burlacu}, {Carballo}, {Casamiquela}, {Chiavassa}, {Cooper},
  {Dapergolas}, {Delchambre}, {Dharmawardena}, {Drimmel}, {Edvardsson},
  {Fouesneau}, {Garabato}, {Garcia-Lario}, {Garcia-Torres}, {Gavel}, {Gomez},
  {Gonzalez-Santamaria}, {Hatzidimitriou}, {Heiter}, {Jean-Antoine Piccolo},
  {Kontizas}, {Korn}, {Lanzafame}, {Lebreton}, {Le Fustec}, {Licata},
  {Lindstrom}, {Livanou}, {Lobel}, {Lorca}, {Magdaleno Romeo}, {Marocco},
  {Marshall}, {Mary}, {Nicolas}, {Pallas-Quintela}, {Panem}, {Pichon},
  {Riclet}, {Robin}, {Rybizki}, {Santovena}, {Silvelo}, {Smart}, {Sarro},
  {Sordo}, {Soubiran}, {Suvege}, {Ulla}, {Vallenari}, {Zorec}, {Utrilla}, \&
  {Bakker}}]{GSPspecDR3}
{Recio-Blanco}, A., {de Laverny}, P., {Palicio}, P.~A., {et~al.} 2022, arXiv
  e-prints, arXiv:2206.05541

\bibitem[{{Recio-Blanco} {et~al.}(2021){Recio-Blanco}, {Fern{\'a}ndez-Alvar},
  {de Laverny}, {Antoja}, {Helmi}, \& {Crida}}]{RecioBlanco2021}
{Recio-Blanco}, A., {Fern{\'a}ndez-Alvar}, E., {de Laverny}, P., {et~al.} 2021,
  \aap, 648, A108

\bibitem[{{Reddy} {et~al.}(2012){Reddy}, {Giridhar}, \& {Lambert}}]{Reddy12}
{Reddy}, A. B.~S., {Giridhar}, S., \& {Lambert}, D.~L. 2012, \mnras, 419, 1350

\bibitem[{{Reddy} {et~al.}(2006){Reddy}, {Lambert}, \& {Allende
  Prieto}}]{Reddy06}
{Reddy}, B.~E., {Lambert}, D.~L., \& {Allende Prieto}, C. 2006, \mnras, 367,
  1329

\bibitem[{{Rizzuti} {et~al.}(2019){Rizzuti}, {Cescutti}, {Matteucci},
  {Chieffi}, {Hirschi}, \& {Limongi}}]{rizzuti2019}
{Rizzuti}, F., {Cescutti}, G., {Matteucci}, F., {et~al.} 2019, \mnras, 489,
  5244

\bibitem[{{Ruiz-Lara} {et~al.}(2020){Ruiz-Lara}, {Gallart}, {Bernard}, \&
  {Cassisi}}]{lara2020}
{Ruiz-Lara}, T., {Gallart}, C., {Bernard}, E.~J., \& {Cassisi}, S. 2020, Nature
  Astronomy, 4, 965

\bibitem[{{Sales-Silva} {et~al.}(2022){Sales-Silva}, {Daflon}, {Cunha},
  {Souto}, {Smith}, {Chiappini}, {Donor}, {Frinchaboy},
  {Garc{\'\i}a-Hern{\'a}ndez}, {Hayes}, {Majewski}, {Masseron}, {Schiavon},
  {Weinberg}, {Beaton}, {Fern{\'a}ndez-Trincado}, {J{\"o}nsson}, {Lane},
  {Minniti}, {Manchado}, {Moni Bidin}, {Nitschelm}, {O'Connell}, \&
  {Villanova}}]{SalesSilva22}
{Sales-Silva}, J.~V., {Daflon}, S., {Cunha}, K., {et~al.} 2022, \apj, 926, 154

\bibitem[{{Santos-Peral} {et~al.}(2020){Santos-Peral}, {Recio-Blanco}, {de
  Laverny}, {Fern{\'a}ndez-Alvar}, \& {Ordenovic}}]{Pablo20}
{Santos-Peral}, P., {Recio-Blanco}, A., {de Laverny}, P.,
  {Fern{\'a}ndez-Alvar}, E., \& {Ordenovic}, C. 2020, \aap, 639, A140

\bibitem[{{Sheffield} {et~al.}(2012){Sheffield}, {Majewski}, {Johnston},
  {Cunha}, {Smith}, {Cheung}, {Hampton}, {David}, {Wagner-Kaiser}, {Johnson},
  {Kaplan}, {Miller}, \& {Patterson}}]{Sheffield12}
{Sheffield}, A.~A., {Majewski}, S.~R., {Johnston}, K.~V., {et~al.} 2012, \apj,
  761, 161

\bibitem[{{Sneden} {et~al.}(2008){Sneden}, {Cowan}, \& {Gallino}}]{sneden2008}
{Sneden}, C., {Cowan}, J.~J., \& {Gallino}, R. 2008, \araa, 46, 241

\bibitem[{{Spina} {et~al.}(2020){Spina}, {Nordlander}, {Casey}, {Bedell},
  {D'Orazi}, {Mel{\'e}ndez}, {Karakas}, {Desidera}, {Baratella}, {Yana
  Galarza}, \& {Casali}}]{Spina20}
{Spina}, L., {Nordlander}, T., {Casey}, A.~R., {et~al.} 2020, \apj, 895, 52

\bibitem[{{Spina} {et~al.}(2021){Spina}, {Ting}, {De Silva}, {Frankel},
  {Sharma}, {Cantat-Gaudin}, {Joyce}, {Stello}, {Karakas}, {Asplund},
  {Nordlander}, {Casagrande}, {D'Orazi}, {Casey}, {Cottrell},
  {Tepper-Garc{\'\i}a}, {Baratella}, {Kos}, {{\v{C}}otar}, {Bland-Hawthorn},
  {Buder}, {Freeman}, {Hayden}, {Lewis}, {Lin}, {Lind}, {Martell},
  {Schlesinger}, {Simpson}, {Zucker}, \& {Zwitter}}]{Spina21}
{Spina}, L., {Ting}, Y.~S., {De Silva}, G.~M., {et~al.} 2021, \mnras, 503, 3279

\bibitem[{{Spitoni} {et~al.}(2022){Spitoni}, {Recio-Blanco}, {de Laverny},
  {Palicio}, {Kordopatis}, {Schultheis}, {Contursi}, {Poggio}, {Romano}, \&
  {Matteucci}}]{spitoni2022}
{Spitoni}, E., {Recio-Blanco}, A., {de Laverny}, P., {et~al.} 2022, arXiv
  e-prints, arXiv:2206.12436

\bibitem[{{Spitoni} {et~al.}(2020){Spitoni}, {Verma}, {Silva Aguirre}, \&
  {Calura}}]{spitoni2020}
{Spitoni}, E., {Verma}, K., {Silva Aguirre}, V., \& {Calura}, F. 2020, \aap,
  635, A58

\bibitem[{{Spitoni} {et~al.}(2021){Spitoni}, {Verma}, {Silva Aguirre},
  {Vincenzo}, {Matteucci}, {Vai{\v{c}}ekauskait{\.{e}}}, {Palla}, {Grisoni}, \&
  {Calura}}]{spitoni2021}
{Spitoni}, E., {Verma}, K., {Silva Aguirre}, V., {et~al.} 2021, \aap, 647, A73

\bibitem[{{Surman} {et~al.}(2008){Surman}, {McLaughlin}, {Ruffert}, {Janka}, \&
  {Hix}}]{Surman08}
{Surman}, R., {McLaughlin}, G.~C., {Ruffert}, M., {Janka}, H.~T., \& {Hix},
  W.~R. 2008, \apjl, 679, L117

\bibitem[{{Tarricq} {et~al.}(2021){Tarricq}, {Soubiran}, {Casamiquela},
  {Cantat-Gaudin}, {Chemin}, {Anders}, {Antoja}, {Romero-G{\'o}mez},
  {Figueras}, {Jordi}, {Bragaglia}, {Balaguer-N{\'u}{\~n}ez}, {Carrera},
  {Castro-Ginard}, {Moitinho}, {Ramos}, \& {Bossini}}]{T21}
{Tarricq}, Y., {Soubiran}, C., {Casamiquela}, L., {et~al.} 2021, \aap, 647, A19

\bibitem[{{Tautvai{\v{s}}ien{\.{e}}} {et~al.}(2021){Tautvai{\v{s}}ien{\.{e}}},
  {Viscasillas V{\'a}zquez}, {Mikolaitis}, {Stonkut{\.{e}}},
  {Minkevi{\v{c}}i{\={u}}t{\.{e}}}, {Drazdauskas}, \& {Bagdonas}}]{Taut21}
{Tautvai{\v{s}}ien{\.{e}}}, G., {Viscasillas V{\'a}zquez}, C., {Mikolaitis},
  {\v{S}}., {et~al.} 2021, \aap, 649, A126

\bibitem[{{Travaglio} {et~al.}(2001){Travaglio}, {Gallino}, {Busso}, \&
  {Gratton}}]{Travaglio01}
{Travaglio}, C., {Gallino}, R., {Busso}, M., \& {Gratton}, R. 2001, \apj, 549,
  346

\bibitem[{{Vincenzo} \& {Kobayashi}(2020)}]{vincenzo2020}
{Vincenzo}, F. \& {Kobayashi}, C. 2020, \mnras, 496, 80

\bibitem[{{Woosley} {et~al.}(1994){Woosley}, {Wilson}, {Mathews}, {Hoffman}, \&
  {Meyer}}]{Woosley94}
{Woosley}, S.~E., {Wilson}, J.~R., {Mathews}, G.~J., {Hoffman}, R.~D., \&
  {Meyer}, B.~S. 1994, \apj, 433, 229

\bibitem[{{Yong} {et~al.}(2008){Yong}, {Karakas}, {Lambert}, {Chieffi}, \&
  {Limongi}}]{Yong08}
{Yong}, D., {Karakas}, A.~I., {Lambert}, D.~L., {Chieffi}, A., \& {Limongi}, M.
  2008, \apj, 689, 1031

\end{thebibliography}

\begin{appendix}
\section{ADQL queries}
\label{Append}
\lstset{language=SQL}

\begin{lstlisting}[caption={\texttt{ADQL} query for the {\it low-uncertainty sample}.} ,captionpos=b]
SELECT source_id
FROM gaiadr3.astrophysical_parameters inner join gaiadr3.gaia_source using(source_id)
WHERE
(rv_expected_sig_to_noise>0)
AND
(vbroad<=13)
AND
(teff_gspspec IS NOT NULL)
AND
(flags_gspspec LIKE '0%') 
AND
(flags_gspspec LIKE '_0%')
AND
(flags_gspspec LIKE '__0%')
AND
(flags_gspspec LIKE '___0%')
AND
(flags_gspspec LIKE '____0%')
AND
(flags_gspspec LIKE '_____0%')
AND
(flags_gspspec LIKE '______0%')
AND
((flags_gspspec LIKE '_______0%') OR (flags_gspspec LIKE '_______1%') ) 
AND
((flags_gspspec LIKE '____________0%') OR ( flags_gspspec LIKE '____________1%') ) 
AND
((flags_gspspec LIKE '%0_____') OR (flags_gspspec LIKE '%1_____') OR (flags_gspspec LIKE '%2_____')) 
AND
((flags_gspspec LIKE '%0____') OR (flags_gspspec LIKE '%1____') ) 
AND
(logchisq_gspspec<-3.75)
AND
(teff_gspspec<=5400)
AND
(logg_gspspec<=3.5)
AND
(cefe_gspspec IS NOT NULL)
AND
( (cefe_gspspec_upper-cefe_gspspec_lower)<=0.4)
\end{lstlisting}

\section{Open cluster data}
\begin{table}
 \caption{Mean of [Fe/H], [Ce/Fe], [Ca/Fe] for our 52 open clusters. }
        \label{Tab:OC}
\begin{tabular}{clcccc}
\hline
    Cluster Name           &   Stars &   [Fe/H] &  [Ce/Fe] &   [Ca/Fe] \\
\hline
 Alessi Teutsch 8  &       1 &     -0.170 &      -0.110 &      -0.140  \\
 Stock 2           &       8 &     -0.131 &      -0.081 &       0.229 \\
 UBC 394           &       1 &     -0.290 &      -0.050 &       0.000  \\
 NGC 2632          &       3 &      0.120 &      -0.133 &       0.177 \\
 Alessi 44         &       1 &     -0.170 &      -0.110 &       0.020  \\
 Roslund 7         &       1 &     -0.200 &      -0.130 &      -0.200   \\
 COIN-Gaia 30      &       1 &     -0.300 &      -0.070 &       0.120  \\
 Trumpler 2        &       2 &     -0.265 &      -0.065 &      -0.070  \\
 UPK 431           &       1 &     -0.190 &      -0.080 &       0.020  \\
 NGC 2281          &       1 &     -0.250 &      -0.070 &       0.150  \\
 IC 2488           &       1 &     -0.230 &      -0.080 &      -0.010  \\
 NGC 5316          &       1 &     -0.210 &      -0.050 &      -0.100   \\
 NGC 2168          &       2 &     -0.140 &      -0.035 &      -0.010  \\
 ASCC 111          &       1 &      0.000 &       0.090 &       0.280  \\
 ASCC 11           &       1 &     -0.300 &      -0.070 &       0.030  \\
 NGC 2682          &       1 &     -0.290 &       0.080 &       0.150  \\
 NGC 5749          &       1 &     -0.290 &      -0.060 &      -0.120  \\
 NGC 2477          &       1 &     -0.360 &       0.060 &       0.070  \\
 NGC 7082          &       1 &     -0.220 &      -0.130 &      -0.170  \\
 NGC 2506          &       1 &     -0.650 &       0.150 &       0.350  \\
 NGC 6633          &       1 &     -0.240 &      -0.060 &       0.100   \\
 UPK 167           &       1 &     -0.150 &      -0.140 &      -0.180  \\
 NGC 2669          &       1 &     -0.180 &      -0.030 &      -0.050  \\
 ASCC 23           &       1 &     -0.190 &      -0.080 &       0.060  \\
 UPK 53            &       1 &     -0.180 &      -0.120 &      -0.060  \\
 Alessi Teutsch 11 &       1 &     -0.190 &      -0.060 &       0.150  \\
 UBC 4             &       1 &     -0.230 &      -0.060 &       0.170  \\
 NGC 1750          &       2 &     -0.245 &      -0.075 &      -0.080  \\
 NGC 1545          &       2 &     -0.300 &      -0.045 &       0.005 \\
 ASCC 71           &       1 &     -0.260 &      -0.110 &      -0.120  \\
 NGC 6475          &       2 &     -0.115 &      -0.090 &      -0.130  \\
 COIN-Gaia 26      &       1 &     -0.180 &      -0.120 &       0.000     \\
 NGC 6124          &       5 &     -0.218 &      -0.094 &      -0.086 \\
 NGC 2447          &       2 &     -0.355 &      -0.015 &       0.005 \\
 NGC 2287          &       2 &     -0.255 &      -0.070 &      -0.030  \\
 NGC 3532          &       3 &     -0.190 &      -0.080 &       0.140  \\
 Stock 1           &       1 &      0.030 &      -0.110 &       0.110  \\
 Collinder 350     &       1 &     -0.290 &      -0.030 &       0.030  \\
 NGC 5662          &       1 &     -0.320 &      -0.060 &       0.000     \\
 NGC 3114          &       3 &     -0.223 &      -0.163 &      -0.037 \\
 UPK 7             &       1 &     -0.220 &      -0.030 &      -0.080  \\
 NGC 6819          &       1 &     -0.350 &       0.180 &       0.500   \\
 Ruprecht 147      &       2 &     -0.075 &      -0.065 &       0.125 \\
 NGC 6281          &       2 &     -0.160 &      -0.065 &       0.070  \\
 NGC 1662          &       1 &     -0.230 &      -0.060 &       0.090  \\
 Platais 8         &       1 &     -0.040 &      -0.090 &       0.280  \\
 UBC 183           &       2 &     -0.295 &      -0.050 &      -0.025 \\
 Gulliver 21       &       1 &     -0.010 &      -0.120 &      -0.230  \\
 IC 4725           &       1 &     -0.050 &       0.030 &      -0.120  \\
 NGC 7789          &       2 &     -0.435 &       0.105 &       0.230  \\
 Collinder 258     &       1 &     -0.300 &      -0.040 &       0.200   \\
 NGC 1647          &       1 &     -0.240 &      -0.100 &      -0.140  \\
 Collinder 463     &       2 &     -0.180 &      -0.125 &      -0.085 \\
\hline
\end{tabular}
\tablefoot{The number of the stars from which we computed these mean values is indicated in the first column.}
\end{table}

\end{appendix}

\end{document}